\documentclass[preprint,authoryear,12pt]{elsarticle}

\usepackage{graphics}

\usepackage{float}
\usepackage{graphicx}
\usepackage{wrapfig}

\usepackage[draft]{todonotes}   

\usepackage[labelfont=bf]{caption}
\usepackage{caption}
\usepackage{color}
\usepackage{amsmath} 
\usepackage{mathtools} 
\usepackage{amsfonts} 
\usepackage{amssymb}
\usepackage{array,longtable}

\usepackage{array}
\usepackage{booktabs}
\usepackage{rotating}
\usepackage{multirow}
\usepackage{multicol}
\usepackage{lscape}
\usepackage{subfigure}

\usepackage[utf8]{inputenc}
\usepackage[T1]{fontenc}
\usepackage{textcomp}
\usepackage{gensymb}

\usepackage[export]{adjustbox}

\usepackage{algorithm,algorithmicx,algpseudocode}

\usepackage{color}

\usepackage[colorlinks=true,linkcolor=blue,citecolor=blue]{hyperref}

\usepackage[draft]{todonotes}   

\usepackage{soul}

\newcommand{\tablefont}{\fontsize{8}{7}\selectfont}

\definecolor{darkgreen}{rgb}{0.53, 0.66, 0.42}
\definecolor{azure}{rgb}{0.0, 0.5, 1.0}
\definecolor{blue}{rgb}{0.0, 0.0, 1.0}
\definecolor{purple}{rgb}{0.55, 0.0, 0.55}
\definecolor{gold}{rgb}{0.72, 0.53, 0.04}
\definecolor{green}{rgb}{0.13, 0.55, 0.13}

\journal{Journal}  

\begin{document}

\begin{frontmatter}



\title{Predicting Infant Brain Connectivity with Federated Multi-Trajectory GNNs using Scarce Data}

\author[first]{Michalis Pistos}
\author[second]{Gang Li}
\author[second]{Weili Lin}
\author[third,forth,fifth]{Dinggang Shen}
\author[first]{Islem Rekik\texorpdfstring{\corref{cor}}}

\cortext[cor]{Corresponding author: i.rekik@imperial.ac.uk: \url{http://basira-lab.com/}.}

\address[first]{BASIRA Lab, Imperial-X and Department of Computing, Imperial College London, London, UK}
\address[second]{Department of Radiology and Biomedical Research Imaging Center, University of North Carolina at Chapel Hill, Chapel Hill, NC, USA} 
\address[third]{School of Biomedical Engineering, ShanghaiTech University, Shanghai 201210, China}    
\address[forth]{Shanghai United Imaging Intelligence Co., Ltd., Shanghai 200230, China}
\address[fifth]{Shanghai Clinical Research and Trial Center, Shanghai, 201210, China}

\begin{abstract}
The understanding of the convoluted evolution of infant brain networks during the first postnatal year is pivotal for identifying the dynamics of early brain connectivity development. Thanks to the valuable insights into the brain's anatomy, existing deep learning frameworks focused on forecasting the brain evolution trajectory from a single baseline observation. While yielding remarkable results, they suffer from three major limitations. First, they lack the ability to generalize to multi-trajectory prediction tasks, where each graph trajectory corresponds to a particular imaging modality or connectivity type (e.g., T1-w MRI). Second, existing models require extensive training datasets to achieve satisfactory performance which are often challenging to obtain. Third, they do not efficiently utilize incomplete time series data. To address these limitations, we introduce FedGmTE-Net++, a \emph{federated graph-based multi-trajectory evolution network}. Using the power of federation, we aggregate local learnings among diverse hospitals with limited datasets. As a result, we enhance the performance of each hospital's local generative model, while preserving data privacy. The three key innovations of FedGmTE-Net++ are: (i) presenting the first federated learning framework specifically designed for brain multi-trajectory evolution prediction in a data-scarce environment, (ii) incorporating an \emph{auxiliary regularizer} in the local objective function to exploit all the longitudinal brain connectivity within the evolution trajectory and maximize data utilization, (iii) introducing a two-step imputation process, comprising a preliminary KNN-based precompletion followed by an \emph{imputation refinement} step that employs regressors to improve similarity scores and refine imputations. Our comprehensive experimental results showed the outperformance of FedGmTE-Net++ in brain multi-trajectory prediction from a single baseline graph in comparison with benchmark methods. Our source code is available at \url{https://github.com/basiralab/FedGmTE-Net-plus}.
\end{abstract}


\end{frontmatter}


\section{Introduction}

Various imaging modalities, such as T1-w MRI and fMRI, offer a rich spectrum of tools to examine the intricate dynamics of brain connectivity. Each modality contributes unique and complementary insights crucial for the diagnosis of neurological disorders. However, acquiring different imaging modalities is challenging, leading to research efforts dedicated to predicting the multi-trajectory evolution of the brain with minimal resources. In particular, a major area of focus in neuroimaging focuses on the infant brain's connectivity development during the first postnatal year, which is a critical period marked by significant structural, morphological, and functional changes \citep{zhang2019resting}. Understanding the infant brain provides insights for early diagnosis of neurodevelopmental diseases, improving the effectiveness of treatment compared to later predictions \citep{stoessl2012neuroimaging}. The ability to predict a predisposition to disease is highly beneficial as it may prevent its occurrence  \citep{rekik2017joint,rekik2016predicting}. Exploring the developmental path of the brain's connectome is crucial for understanding its dynamic nature. Consequently, numerous studies have emerged to forecast the longitudinal evolutionary trajectory of the brain. Early studies including \citep{ghribi2021multi} and \citep{ezzine2019learning} introduced simple frameworks that adopt a sample selection strategy, utilizing K-Nearest Neighbours (KNN) \citep{peterson2009k} for the final predictions. Although promising, several attempts of more advanced deep learning methods were made in an attempt to achieve better performance.

The brain connectome can be represented as a graph, where nodes represent distinct regions of interest (ROIs) \citep{liu2011few}. The number of ROIs can vary based on MRI parcellation techniques like AAL or MNI templates \citep{tzourio2002automated}. Graph edges encode the strength of connections between ROI pairs. Graph Neural Networks (GNNs) \citep{velivckovic2023everything, zhou2020graph, bessadok2022graph} have already demonstrated noteworthy success in various biomedical studies, including disease classification \citep{rhee2017hybrid, parisot2018disease} and protein interaction prediction \citep{gainza2020deciphering, fout2017protein}. Hence, they are increasingly becoming a prevalent foundation for deep learning frameworks in tasks related to brain network prediction.
Existing studies \citep{goktacs2020residual, hong2019longitudinal, nebli2020deep, tekin2021recurrent, demirbilek2023predicting} propose novel GNN architectures, including Graph Convolutional Networks (GCNs) \citep{kipf2016semi}, graph Generative Adversarial Networks (gGANs) \citep{wang2018graphgan}, and Graph Recurrent Networks (GRNs) \citep{liao2019efficient, medsker2001recurrent} in their brain trajectory prediction task.

\begin{figure}[ht!]
  \centering
  \includegraphics[width=6in, height=2.5in]{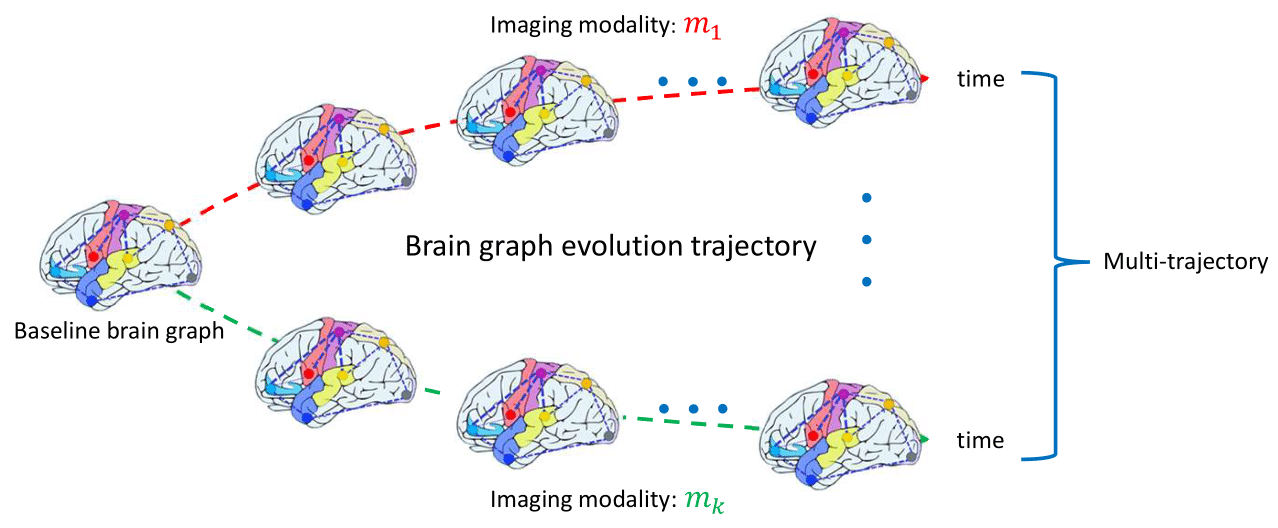}
  \caption{Brain graph multi-trajectory evolution including different imaging modalities with a single baseline brain graph as input. We span the full multimodal space of 4D trajectories from a single brain connectome.}
  \label{fig:multi_trajectory}
\end{figure}

While effective, these methods typically require extensive training data for satisfactory performance and generalization \citep{hestness2017deep}. However, obtaining medical data is challenging due to long acquisition times, high costs, and limited subject availability. Current methods are also restricted to single-trajectory prediction, but generalizing to multi-trajectory prediction (see \textbf{Fig.} \ref{fig:multi_trajectory}) enables the generation of diverse brain graphs from various imaging modalities starting from a baseline graph (e.g., functional and morphological trajectories), ideally utilizing the least costly to acquire (i.e., affordable). The integration of brain graphs derived from multiple imaging modalities offers complementary insights into the brain dynamics and its multi-facet connectivity, leading to improved diagnostic accuracy (i.e., Alzheimer's disease diagnosis \citep{meng2022multi}). Hence, a pressing need exists for a multi-trajectory predictive model that tackles \emph{data scarcity}. Only one paper in the existing literature \citep{bessadok2021few} proposed a few-shot learning framework for multi-trajectory evolution using a Teacher-Student (TS) learning paradigm \citep{hinton2015distilling}. Despite successfully augmenting the training dataset with the use of simulated data, the TS paradigm has downsides, including increased computational costs, noise amplification, and a risk of bias and poor generalization.

We aim to present a robust framework that addresses the limitations inherent in existing methods. Thus, our paper explores the concept of federated learning, a paradigm that leverages multiple locally small datasets to achieve better performance without additional computational costs or poor generalization drawbacks.
Federated learning has gained popularity for enabling the collaboration among independent decentralized clients, such as hospitals, to improve model training without compromising the privacy of sensitive information. By combining local learnings, models can be trained on diverse datasets, enhancing generalization and mitigating biases. While recent works \citep{gurler2020foreseeing, gurler2022federated} have introduced the first federated brain graph evolution trajectory prediction frameworks, the proposed 4D-Fed-GNN framework and its variants are only applicable to unimodal connectomic datasets. In this paper, we first present Federated Graph Multi-Trajectory Evolution Network (FedGmTE-Net), which forms our foundational model, employing federated learning to address data limitations. Two refined variants, FedGmTE-Net+ and FedGmTE-Net++, are later introduced. FedGmTE-Net+ utilizes a KNN-based imputation technique based on baseline observations to fill in the incomplete training data and incorporates an \emph{auxiliary regularizer} for optimal utilization of all available longitudinal brain graphs. FedGmTE-Net++ improves on the naive imputation performed in FedGmTE-Net+ by employing an \emph{imputation refinement} step post-initial training. This step leverages the model's predictions to train regressors with the aim of refining the initial imputations and fine-tuning our model's performance. In summary, our main contributions are as follows:
\begin{enumerate}
    \item We propose Federated Graph Multi-Trajectory Evolution Network (FedGmTE-Net) which is the first graph multi-trajectory framework that leverages federation for forecasting the infant brain evolution trajectory.
    \item We upgrade the standard MAE objective function by including a topology loss to preserve the graph's topological patterns.
    \item We introduce FedGmTE-Net+, which incorporates a novel auxiliary regularizer to enhance data utilization in a limited data environment. To address missing data in longitudinal graph datasets, it employs a KNN-based imputation technique based on baseline observations.
    \item We further propose FedGmTE-Net++, which after the first training round employs modality and hospital-specific similarity regressors to improve the initial imputations and subsequently conducts a few rounds of fine-tuning to achieve better results.
\end{enumerate}

Note that a preliminary version of this work (FedGmTE-Net*) was published in \citep{pistos2023federated}. This journal version presents the following extensions. (1) To optimize the utilization of incomplete subjects and improve on the simplistic imputation approach of FedGmTE-Net+, we introduce FedGmTE-Net++, which incorporates an additional imputation refinement step. This involves training \emph{distinct regressors} for each imaging modality and each hospital to enhance the initial imputations utilized as ground-truth for missing data, thereby elevating the overall predictive capabilities of the network. (2) To better showcase the effectiveness of our method in diverse data scenarios we conducted additional experiments showcasing results across IID, non-IID, real and simulated datasets. (3) In order to illustrate the enhanced performance of the newly proposed framework, we conducted a thorough comparison with its prior architectures. (4) For a more comprehensive evaluation, we added new evaluation measures including the Pearson Correlation Coefficient (PCC) and Jaccard distance (JD).

\section{Related Work}
\textbf{Brain evolution prediction.} Several studies have explored diverse machine-learning methods for forecasting the evolutionary trajectory of the brain network.  A notable comparative study by \citep{akti2022comparative} benchmarks 20 cutting-edge machine learning pipelines tailored for forecasting temporal changes in the brain connectome. Recent studies on brain graph trajectory prediction can generally be categorized into two groups: those employing prediction with a sample selection strategy, and those leveraging advanced deep learning frameworks, mostly based on GNNs. A noteworthy example in the former category is \citep{ghribi2021multi}, which uses three distinct bidirectional regressors to compute similarity scores between subjects in supervised multi-regression. Alternatively, \citep{ezzine2019learning} proposed the Learning-guided Infinite Network Atlas selection (LINAs) framework, utilizing unsupervised learning and an infinite sample selection strategy to determine pairwise similarities between subjects. 
Despite accurately predicting brain network evolution, these models were divided into subparts during learning, lacking an end-to-end learning fashion. 

To address this, advanced deep-learning methods have been proposed in the literature.
Given the capacity to represent the brain through graphs, employing GNNs as a foundational framework for predictive models has become widely adopted. For instance, \citep{goktacs2020residual} designed RESNets which integrated GCNs to facilitate the learning of an adversarial brain graph embedding, which was then used in a sample selection strategy guided by connectional brain templates (CBTs). Similarly, \citep{hong2019longitudinal} employs GCNs for predicting missing infant brain trajectory data through adversarial learning. Further, \citep{nebli2020deep} introduced EvoGraphNet, employing time-dependent gGANs, while \citep{tekin2021recurrent} presented Recurrent Brain Graph Mapper (RBGM), using recurrent graph neural networks instead. Finally, in \citep{demirbilek2023predicting}, the proposed ReMI-Net, identifies biomarkers that differentiate between healthy and disordered populations by predicting the evolution of population-driven CBTs. 
The aforementioned papers share two notable limitations: the need for a large training dataset for improved model learning and a lack of emphasis on multi-trajectory prediction. \citep{bessadok2021few} tackles these issues by introducing a novel few-shot learning framework for multi-trajectory evolution. Using a Teacher-Student (TS) paradigm, which is particularly well-suited for few-shot learning contexts, two distinct networks, teacher and student, are learned. The teacher is trained on the original dataset, while the student is trained on simulated data mirroring the original dataset's distribution. The student leverages simulated data to enrich its training set and benefits from the knowledge of the teacher network, trained on real data. However, the TS framework has drawbacks, including increased computational expenses both in time and storage due to the use of two separate networks. Additionally, the heavy reliance of the student on the teacher network introduces potential risks of noise and bias, hindering its effective generalization.

\textbf{Federated learning.} Federation \citep{mcmahan2017communication, li2020federated, smith2017federated}, emerges as a well-suited paradigm for decentralized and distributed learning from diverse datasets. This approach provides a robust solution for addressing challenges posed by limited data availability. It involves collaboratively training models on independent data sources while prioritizing data privacy. Unlike traditional centralized methods, federated learning allows individual clients to keep their data locally and only share model updates with a central server serving as the coordinator. Thus, it provides an ideal and secure framework for healthcare institutions, fostering collaboration in data-scarce scenarios. The widely adopted Federated Averaging (FedAvg) \citep{mcmahan2017communication} stands as a renowned federated learning paradigm, serving as inspiration for various works like \citep{li2020fed_prox, acar2021federated}. The algorithm operates in the following manner:  In a scenario with $K$ clients (in our case hospitals), a subset $C \in (0, 1]$ is chosen for local training across $E$ epochs. Following local training, client devices send their model updates to a central server, which aggregates them through averaging to produce a new global model. This iterative process ensures the incorporation of knowledge from all participating devices, benefiting the overall model performance across the federated network. 
Recently, \citep{gurler2020foreseeing, gurler2022federated} introduced 4D-Fed-GNN and its variants for federated brain graph evolution trajectory prediction. The framework addresses data quality challenges in hospitals and aims to improve predictions for hospitals with data gaps by leveraging knowledge from those with complete data. The paper proposed a dual federation strategy, involving an initial aggregation followed by a weight exchange step. The exchange is performed based on a sophisticated time-dependent ordering, organizing hospitals into a daisy chain based on data availability. However, the paper has limitations, such as being restricted to unimodal connectomic datasets and assuming identical data availability among subjects from the same hospital. Additionally, the approach requires separate models for each timepoint of interest, leading to significant storage overhead.

\section{Method}

\begin{figure*}
\centering
{\includegraphics[width=13.5cm]{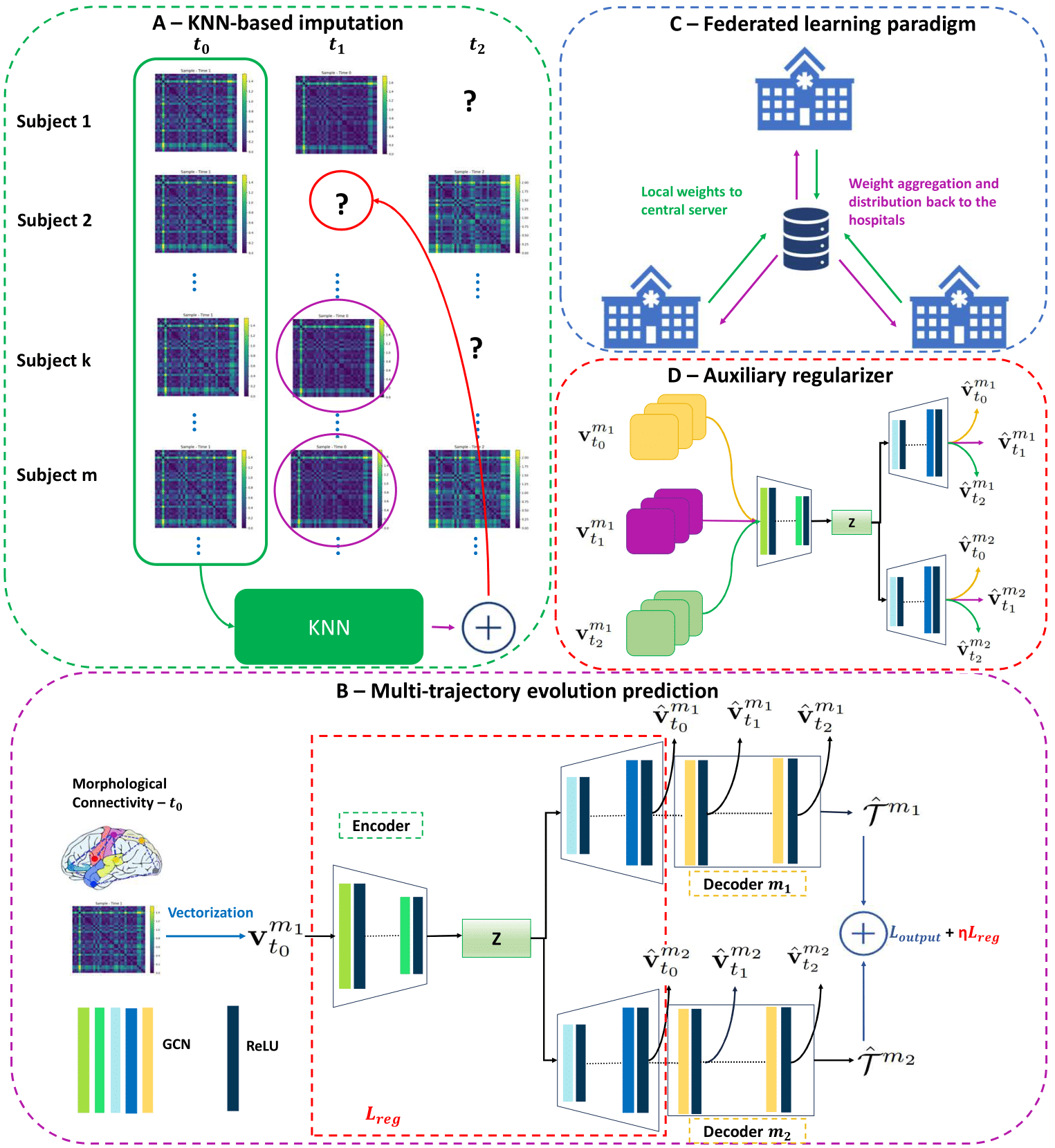}}
\caption{\emph{Pipeline of proposed FedGmTE-Net+ for infant brain graph multi-trajectory evolution prediction from baseline.} \textbf{(A) KNN-based imputation} Imputation technique to complete the missing graphs from each hospital's local training set by utilizing the similarities of subjects at the baseline timepoint. \textbf{(B) Multi-trajectory evolution network} Each hospital's network uses a single input modality to generate multiple trajectories spanning different imaging modalities. \textbf{(C) Federated learning paradigm} A decentralized learning paradigm that allows hospitals to collaborate with each other without sacrificing data privacy. \textbf{(D) Auxiliary regularizer} The auxiliary regularizer improves network performance by utilizing the entire local training dataset across all timepoints.}
\label{fig:main_fed+}
\end{figure*}

\begin{figure*}[ht!]
\centering
\includegraphics[width=\textwidth]{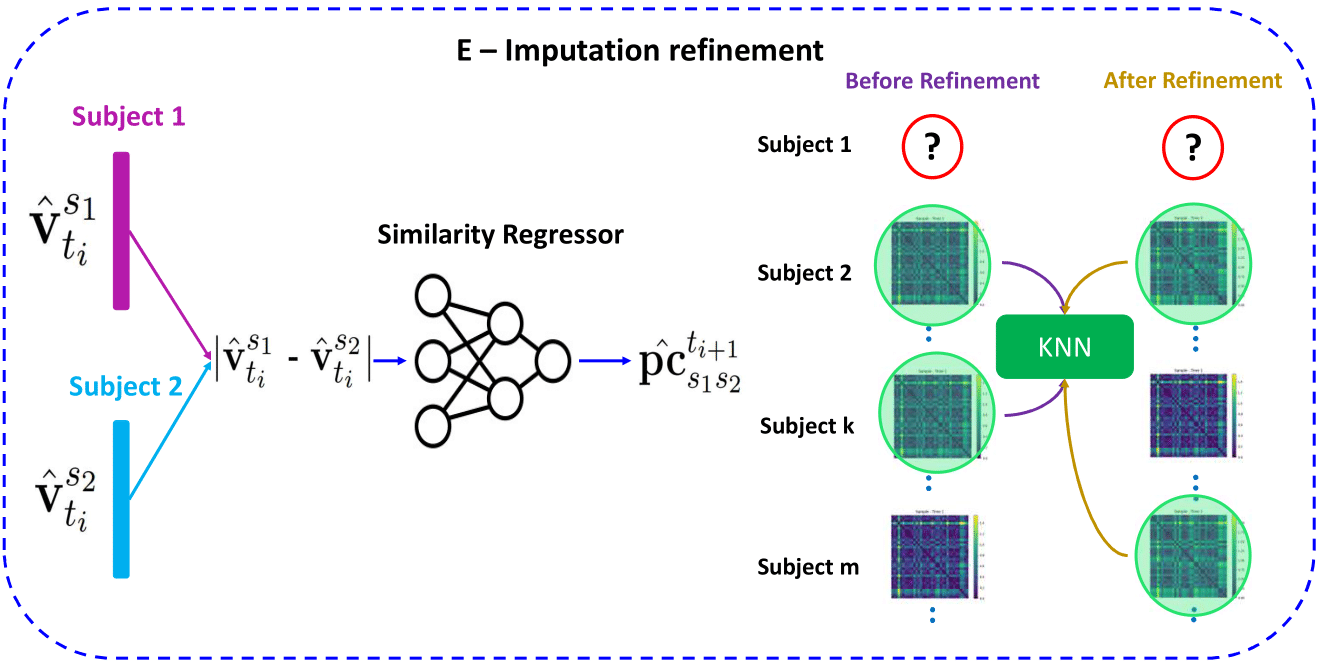}
\caption{\textbf{(E) Imputation refinement step} for the proposed FedGmTE-Net++ network. Following the initial training of the network, we proceed to train the similarity regressors. Their purpose is to update the similarity scores for subject pairs across various timepoints. The updated scores are used to improve the nearest neighbours' selection for each subject, consequently refining the initial KNN imputation. Subsequent to this step, a series of fine-tuning rounds are executed, capitalizing on the improved dataset.}
\label{fig:main_fed++}
\end{figure*}

In our work, we introduce FedGmTE-Net++ and its predecessor models, which are frameworks specifically designed for forecasting the multi-trajectory evolution of a baseline brain graph (i.e., one connectivity type) using federated learning. \textbf{Fig.} \ref{fig:main_fed+} and \textbf{Fig.} \ref{fig:main_fed++} provide an overview of our proposed method,  while Table \ref{notations} provides a summary of the key mathematical notations used in this paper.

\begin{table}[ht!]
    \small
    \centering
    \begin{tabular}{cp{11cm}}
    \hline Notation & Definition \\
    \hline $\mathcal{G}$ & graph representing the brain connectome
    \\
    $N$ & number of graph nodes (ROIs)
    \\
    $d$ & node feature vector dimensionality
    \\
    $\mathbf{X}$ & graph nodes (ROIs) feature matrix $\in \mathbf{R}^{N \times d}$
    \\
    $\mathbf{A}$ & graph adjacency matrix $\in \mathbf{R}^{N \times N}$
    \\
    $\mathbf{v}$ & reduced brain fecture vector from graph
    \\
    \textasciicircum & model predicted values
    \\
    $n_{s}$ & total number of training subjects across hospitals
    \\
    $s$ & subject index
    \\
    $n_{m}$ & number of different imaging modalities
    \\
    $m_{j}$ & $j^{th}$ imaging modality in the network
    \\
    $n_{h}$ & number of hospitals
    \\
    $n_{s}^{h}$ & number of training subjects at hospital h
    \\
    $\{\mathcal{T}_{s}^{m_{j}}\}$ & graph trajectory for subject $s$ and imaging modality $m_{j}$
    \\
    $n_{t}$ & number of timepoints in evolution trajectory
    \\
    $t_{i}$ & $i^{th}$ timepoint of interest
    \\
    $\mathbf{v}_{t_{i}}^{m_{j},s}$ & brain graph feature vectors for timepoint $t_{i}$, modality $m_{j}$ and subject $s$
    \\
    $\mathbf{v}_{t_{i}}^{m_{j}}$ & brain graph feature vectors for timepoint $t_{i}$ and modality $m_{j}$ (all subjects included)
    \\
    $\mathbf{v}_{t_{i}}^{s}$ & brain graph feature vectors for timepoint $t_{i}$ and subject $s$ (all modalities included)
    \\
    $D^{m_{j}}$ & set of decoder networks for modality $m_{j}$
    \\
    $\mathbf{N}_{t_{i}}^{m_{j},s}$ & node strength for subject $s$ at timepoint $t{i}$ and modality $m_{j}$
    \\
    $\mathbf{N}_{r_{i}}$ & node strength for ROI $r_{i}$
    \\
    $\mathcal{L}_{L1}$ & L1 (graph MAE) loss term
    \\
    $\mathcal{L}_{tp}$ & topology loss term
    \\
    $\lambda$ & topology loss term regularizer
    \\
    $\mathcal{L}_{output}$ & output loss: $\mathcal{L}_{output}= \mathcal{L}_{L1} + \lambda \mathcal{L}_{tp}$
    \\
    $L_{reg}$ & auxiliary loss term
    \\
    $\eta$ & auxiliary loss term regularizer
    \\
    $\mathcal{L}_{total}$ & total loss: $\mathcal{L}_{total}= \mathcal{L}_{output} + \eta\mathcal{L}_{reg}$
    \\
    $\mathbf{pc}_{s_{1}s_{2}}^{t_{i}}$ & Pearson correlation coefficient between subject $s_{1}$ and subject $s_{2}$ at timepoint $t_{i}$
    \\
    \hline
    \end{tabular}
\caption{Major mathematical notations.}
\label{notations}
\end{table}

\textbf{Problem defintion.} A brain connectome can be represented as a graph denoted by $\mathcal{G}=\{\mathbf{X}, \mathbf{A}\}$, where $\mathbf{X}$ denotes the different ROIs of the brain. Each ROI acts as a node in the graph, resulting in the construction of the matrix $\mathbf{X} \in \mathbf{R}^{N \times d}$, with $N$ distinct ROIs, each linked to its unique $d$-dimensional feature vector. The brain graph adjacency matrix, denoted as $\mathbf{A} \in \mathbf{R}^{N \times N}$ is a weighted matrix that represents the connectivity strength between different brain regions (nodes of the graph). In our framework, we vectorize each brain graph $\mathcal{G}$ into a reduced feature vector, referred to as $\mathbf{v}$, containing a summary of the entire connectivity weights in our graph. Specifically, given that a brain graph is represented by a symmetric matrix $\mathbf{X}$, we extract its feature vector by vectorizing the off-diagonal upper-triangular part of $\mathbf{X}$. Since our problem involves the evolution prediction of multiple trajectories from a single baseline graph, each of our subjects $s \in \{1,..., n_{s}^{h}\}$ consists of multiple graph trajectories derived from different imaging modalities: $\{\mathcal{T}_{s}^{m_{j}}\}_{j=1}^{n_{m}}$, where $n_{s}^{h}$ denotes the number of training subjects in arbitrary hospital $h$ and $n_{m}$ stands for the number of modalities. The graph trajectory for a specific modality $m_{j}$ of a subject can be expressed as $\mathcal{T}_{s}^{m_{j}}= \{\mathbf{v}_{t_{i}}^{m_{j},s}\}_{i=1}^{n_{t}}$, which includes the respective brain graph feature vectors at all $n_{t}$ timepoints. Our GNN takes in a single graph derived from a conventional MRI modality (the most trivial to acquire by all hospitals, e.g., T1-w) at the baseline timepoint $t_0$, represented as $\mathbf{v}_{t_{0}}^{m_{1}}$. The desired output is the prediction of the trajectory evolution across all modalities and timepoints, given by the multi-trajectories $\{\{\hat{\mathbf{v}}_{t_{i}}^{m_{j}}\}_{i=1}^{n_{t}}\}_{j=1}^{n_{m}}$.

\begin{figure}[ht!]
  \centering
  \includegraphics[width=4.8in, height=2.7in]{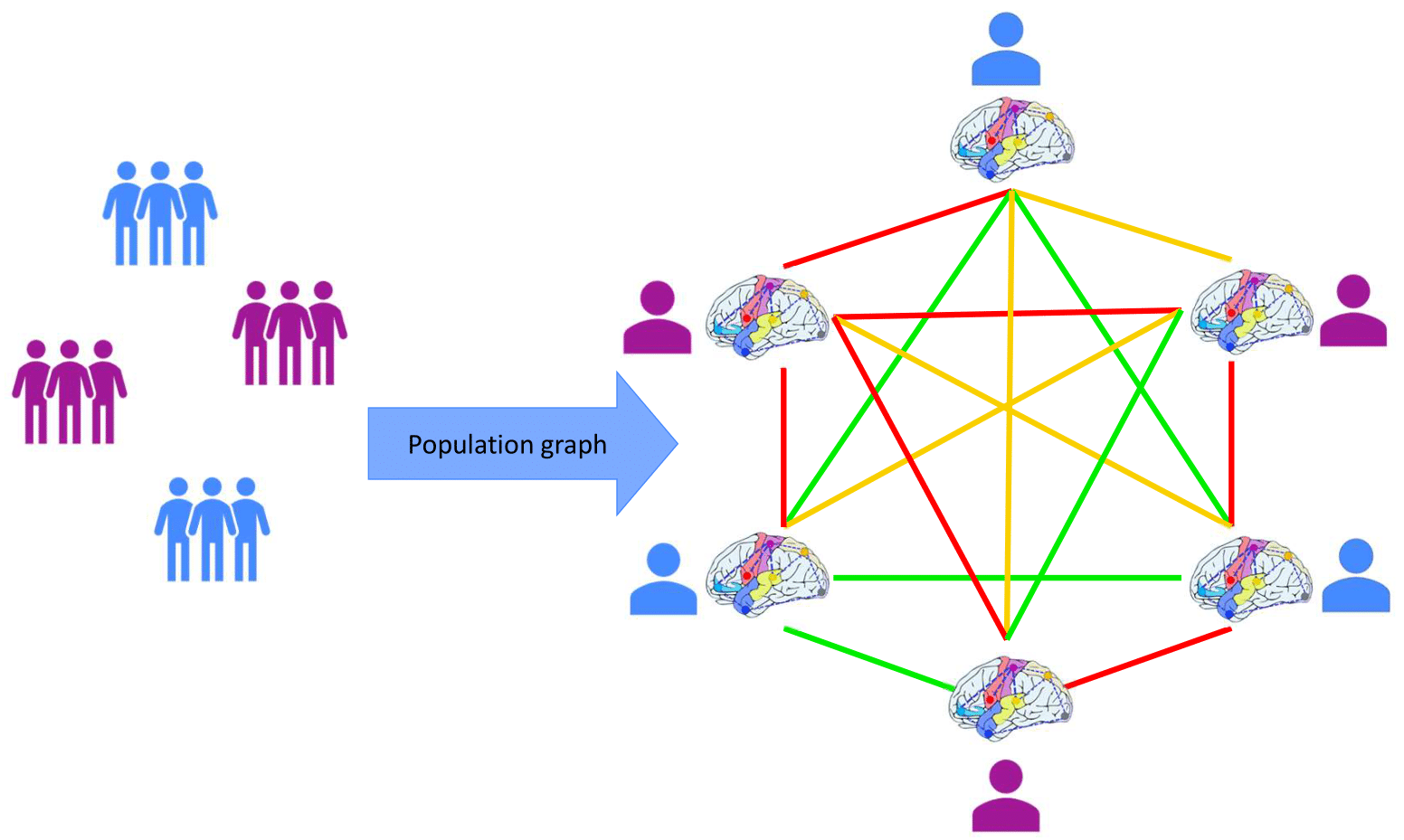}
  \caption{Hospital-specific population graph where nodes symbolize distinct subject brain graphs within the population, and edges denote their connections. The use of various colours for the connections showcases the non-binary nature of the relationship between two subjects.}
  \label{fig:pop_graph}
\end{figure}

\textbf{Population-based graphs.} As noted earlier, the number of ROIs in the brain corresponds to the number of graph nodes, which can vary across different imaging modalities. Preserving the original graph structure becomes challenging since we are dealing with different numbers of nodes in a GNN, hence the vectorization step. However, we want to take advantage of the benefits of GNNs. Thus, we construct a  unique population-based graph for each hospital, including its entire subject population. By doing so, we get a comprehensive representation of the relationships and connections between subjects. Consequently, subjects can learn from others who possess similar structures. An example representation of a population-based graph, which includes the entire population can be found in \textbf{Fig.} \ref{fig:pop_graph}. In this context, the nodes correspond to the subjects, whose graphs are transformed into vectors, constituting the features of their respective nodes. The edges of the population graph symbolize the connections that reflect their similarities. The similarity connections are not confined to a binary distinction determining whether two subjects are similar or not, but rather take the form of scalar weights that can differ from one subject pair to another. There are various methods available to compute the similarity between subjects, which will be denoted by the edges of the graph. In our framework, we measure the dissimilarity between two subjects by calculating the Euclidean distance between their feature vectors, followed by applying the exponential function to generate an adjacency matrix of the graph population per hospital.

\subsection{FedGmTE-Net+}
Extending our prime model FedGmTE-Net which simply integrated the federated learning concept into our multi-trajectory prediction framework, we first introduce FedGmTE-Net+ incorporating several innovative methodologies designed to address data scarcity issues. The pipeline of the proposed FedGmTE-Net+ can be found in \textbf{Fig.} \ref{fig:main_fed+}.

\textbf{A - KNN-based imputation.} Longitudinal medical datasets are often incomplete. Our framework adopts a subject-level availability assumption, which allows for different subjects to have \emph{varying} acquisition timepoints. Compared to other feration paradigms such as hospital-level availability \citep{gurler2020foreseeing, gurler2022federated}, where subjects from the same hospitals have identical availability, this approach is much more realistic by acknowledging the potential variability across subjects. Further, we assume that the baseline timepoint ($t_{0}$) should be available for all subjects. Due to the limited nature of our dataset, we cannot afford to exclude subjects with missing values. Therefore, we employ a KNN-based imputation technique to utilize all available subject graphs for each hospital independently. The algorithm works as follows. When a subject has missing graphs at a specific timepoint, we identify its closest neighbours with available graphs at that timepoint, using the Pearson correlation coefficient. To impute the missing graphs, we calculate the average of the brain graphs from these neighbours. As all subjects have complete data at the baseline timepoint, we identify the nearest neighbours based on their graph similarities at $t_{0}$. The underlying idea is that if subjects are similar at baseline, they should exhibit similar patterns over follow-up timepoints. This imputation procedure is applied independently for each available modality in the dataset and for each local hospital. An example of the described imputation technique for baseline graphs can be found in \textbf{Fig.} \ref{fig:main_fed+}-A.

\textbf{B - Graph multi-trajectory evolution prediction.} The network architecture utilized in this work is inspired by \citep{bessadok2021few}. For each hospital involved in our federated learning framework, we adopt the same autoencoder architecture. This architecture incorporates GCN layers to construct highly
expressive representations \citep{wu2020graph}, by utilizing local connections and shared weights. Initially, an encoder is employed to learn a meaningful latent representation using the input modality at $t_{0}$. The encoder module is composed of a two-layer GCN, where each layer is followed by a ReLU activation and a dropout function. Subsequently, a set of decoder networks, denoted as $\{D^{m_{j}}\}_{j=1}^{n_{m}}$, is used for each output graph from modality $m_j$. These decoders take the representation generated by the encoder as input. To account for variations in graph resolution (i.e., node size) between the initial and target modalities and subsequently evolution trajectories, the first decoding step involves a rescaling process to match the dimensions of the corresponding output modality graph. Thus, our decoder is capable of predicting brain graphs even when resolution shifts occur across different graphs. The rescaling module of the decoder follows the same two-layer GCN architecture as the encoder. Finally, following this, an elementary GCN-based module is cascaded $n_{t} - 1$ times, where each iteration predicts a brain graph matrix at a distinct timepoint. 

Our primary prediction loss for a hospital $h$ consists of an $L1$ loss and a topological loss term $L_{tp}$:

$$
\mathcal{L}_{output}= \mathcal{L}_{L1}\left(\hat{\mathbf{v}}, \mathbf{v}\right) + 
\lambda \mathcal{L}_{tp}\left(\hat{\mathbf{N}}, \mathbf{N}\right)
$$
$$
\mathcal{L}_{L1}\left(\hat{\mathbf{v}}, \mathbf{v}\right)=\frac{1}{n_m n_t n_s^h} \sum_{j=1}^{n_m} \sum_{i=1}^{n_t} \sum_{s=1}^{n_s^h} \lvert{\hat{\mathbf{v}}_{t_{i}}^{m_{j},s}}-{\mathbf{v}}_{t_{i}}^{m_{j},s}\rvert 
$$
$$
\mathcal{L}_{t p}\left(\hat{\mathbf{N}}, \mathbf{N}\right)=\frac{1}{n_m n_t n_s^h} \sum_{j=1}^{n_m} \sum_{i=1}^{n_t} \sum_{s=1}^{n_s^h} \left(\hat{\mathbf{N}}_{t_{i}}^{m_{j},s}-\mathbf{N}_{t_{i}}^{m_{j},s}\right)^{2}
$$

The $L1$ loss measures the difference between the predicted and actual brain graph feature vectors. It quantifies the mean absolute error (MAE) between the predicted and ground-truth brain graphs across all modalities. To retain the unique topological properties of brain connectomes and minimize topological dissimilarities between the predicted and ground-truth brain graphs, a topological loss is incorporated, with the node strength \citep{newman2004analysis} as a topological measure. The topological strength of each node is computed by summing all its edge weights. Next, the topological loss is calculated as the mean squared error (MSE) between the true and predicted graphs. The topological loss is tuned using the hyperparameter $\lambda$. Finally, the output loss ($L_{output}$) described above is combined with an auxiliary regularizer loss ($L_{reg}$), giving the total loss $\mathcal{L}_{total}= \mathcal{L}_{output} + \eta\mathcal{L}_{reg}$. The auxiliary loss term is also adjusted by a hyperparameter $\eta$, and will be introduced below. By employing a weighted multi-loss function, we can manage the impact of each loss term on the overall optimization process.

\textbf{C - Federated learning paradigm.} In our case, the local medical datasets of individual hospitals are insufficient to train a generalizable model. Hence, we adopt a federated learning framework, which enables distributed learning. This approach allows hospitals to train their models collaboratively while preserving the data privacy of their subjects and hence improving each hospital's individual model, even when trained on a small dataset. In particular, we use FedAvg introduced in \citep{mcmahan2017communication}, which works with $n_{h}$ hospitals and a shared global server. At the start of each global federation round, the hospitals receive the current global state of the shared model and undergo local training on their respective datasets. Once the local training is finished, the hospitals send their respective model updates back to the central server/coordinator. The server aggregates these updates using a weighted average, resulting in a new global model. This averaging process ensures that the knowledge from all participating hospitals is incorporated into the global model, benefiting everyone. Further, using a weighted average as the aggregation strategy accommodates hospitals with different dataset sizes by assigning a higher weight to those with more subjects as they are more representative. The aggregation strategy is implemented throughout the entire network, which is outlined in the preceding section. Specifically, it involves all hospital-specific weights ($w^h$) learned by the GCN layers within both the encoder and decoder models which are then combined to form the weights of the global model ($w$). The aggregation strategy is mathematically expressed by the equation below:

$$
w_{t+1} \leftarrow \sum_{h=1}^{n_{h}} \frac{n_{s}^{h}}{n_{s}} w_{t+1}^{h}
$$

One of the primary challenges in this training paradigm is the high communication cost. To mitigate this, we make a trade-off with computation costs, by performing five local training rounds at each hospital instead of one before sending the updates to the central server. We avoid using a higher number of local training epochs since this can result in divergence issues due to significant differences in clients' parameters.

\textbf{D - Auxiliary regularizer.} To increase the robustness and accuracy of our training models, we further incorporate an auxiliary loss term at each hospital's local objective function. The goal of the new term is to better utilize our available longitudinal dataset. The term introduced is defined as $L_{reg}$ and specifically aims at improving the performance of our network encoder and the rescaling process performed by the decoders, which is the module surrounded by a dotted red border in \textbf{Fig.} \ref{fig:main_fed+}-B. Currently, we only train this using brain graphs from timepoint $t_{0}$. However, this module does not seem to rely on the temporal nature of our data, as it simply encodes the input baseline graph into a latent representation and then rescales it to the target modalities. During this process, we do not move forward in time, since the input and target modalities share the same timepoint. Therefore, we can instead utilize all available data, including brain graphs at all timepoints ($t_{0}$, $t_{1}$, $t_{2}$, ..) for our module training. This is precisely what the $L_{reg}$ term addresses, which calculates the loss of the module across all available timepoints (see \textbf{Fig.} \ref{fig:main_fed+}-D). This loss is calculated in the same fashion as the output loss but utilizes predictions solely based on the rescaling module, as opposed to the entire network. Subsequently, the computed $L_{reg}$ is combined with the standard output loss $L_{output}$ to give the final $L_{total}$. By improving the performance of this module first, our goal is to enhance the initial prediction of our local hospital's network at the baseline timepoint ($t_{0}$). This raises the question of how subsequent timepoints, such as $t_{1}$ and $t_{2}$, are affected. Given the nature of time-series prediction scenarios, we anticipate an overall improvement in performance across subsequent timepoints too. The underlying assumption is that a more accurate prediction at a previous timepoint, which is closer to the ground-truth, would also lead to improved predictions at the subsequent timepoints. Consequently, our improvement at $t_{0}$ is expected to initiate a chain of events that ultimately improves performance across all timepoints.

\subsection{FedGmTE-Net++}
The second iteration of FedGmTE-Net, known as FedGmTE-Net++, is introduced with the aim of improving the existing FedGmTE-Net+ framework. It addresses a key limitation of FedGmTE-Net+, specifically its simplistic imputation approach, by introducing an imputation refinement step (see \textbf{Fig.} \ref{fig:main_fed++}).

\textbf{E - Imputation refinement}
The motivation for refining imputations originated from the limitations of the initial imputation process used in FedGmTE-Net+, which is overly simplistic and not optimal. As discussed earlier, FedGmTE-Net+'s imputation method relies on assumptions derived from comparing subjects solely based on their baseline observations since no other information is available during the initial data precompletion prior to training. However, in FedGmTE-Net++, we introduce a refinement step to enhance imputations after the initial training, aiming to achieving better performance. We have confidence that our model, after extensive training, excels in generating predictions across subjects and timepoints. Therefore, we aim to leverage these predictions, instead of relying solely on the baseline timepoint from the ground-truth data, to update the similarity scores between local hospital subjects, thereby improving the imputations. The concept of this innovative framework is outlined as follows:
\begin{enumerate}
    \item \textbf{Initial Training.} Initially, we follow the standard training procedure as described above, following the same steps as we do in the FedGmTE-Net+ method.
    \item \textbf{Train Similarity Regressors.} Upon completing the training process, the subsequent step involves training the similarity regressors. These regressors constitute the central component of the final imputation refinement step and their goal is to update the similarity scores between subjects. There are distinct regressors for each modality and each hospital, yet they are shared across different timepoints. Their architecture is a simple neural network comprising two hidden layers with ReLU activation, followed by one output layer with sigmoid activation. The input to the regressors consists of the absolute difference between the predictions made by our network for two distinct subjects ($s_{1}$ and $s_{2}$) at timepoint $t_{i}$, denoted as $\hat{\mathbf{v}}_{t_{i}}^{s_{1}}$ and $\hat{\mathbf{v}}_{t_{i}}^{s_{2}}$ respectively. Utilizing their absolute prediction difference, the regressors forecast the positive Pearson correlation coefficient of the ground-truth data at the subsequent timepoint $t_{i+1}$, which is expressed as $\hat{\mathbf{pc}}_{s_{1}s_{2}}^{t_{i+1}}$. In this way, we leverage our network's predictions at a timepoint to estimate the similarity between subjects at the next timepoint. To train the similarity regressors effectively, we utilize all available tuples of ($s_{1}$, $s_{2}$, $t_{i+1}$) for which we possess ground-truth training data to calculate the expected Pearson correlation coefficient.
    \item \textbf{Update Imputations.} Once we have trained the similarity regressors, we can employ them to improve the initial similarity scores between subjects. In particular, all the similarity scores among all tuples ($s_{1}$, $s_{2}$, $t_{i}$) are updated. Subsequently, for each subject and timepoint that initially lacked data, we can identify its new closest neighbours based on the highest predicted similarity scores and aggregate their brain graphs to receive the final and refined imputation.
    \item \textbf{Fine-tuning.} Finally, we proceed to retrain our network for a few epochs, utilizing the refined dataset. This is a form of fine-tuning, where just a few rounds of additional training are performed on the originally trained network to further elevate its performance.
\end{enumerate}

\section{Results}

\subsection{Evaluation datasets}

\textbf{Real dataset.} We evaluate our proposed framework using the longitudinal multimodal infant brain connectomes, obtained from an in-house dataset. Each subject in the dataset has serial resting-state fMRI, as well as T1-w and T2-w MRI acquired at different timepoints. Using the above data, two types of brain modalities are generated for each subject. Firstly, after rsfMRI pre-processing (including motion
correction), we performed infant brain image longitudinal registration from native space to MNI space using GLIRT where each rsfMRI was partitioned into 116 distinct ROIs with the use of an AAL template. Hence, a 116$\times$116 connectivity matrix is created for each subject, where the connectivity strength between the ROIs corresponds to the Pearson correlation between their mean functional signals. In addition to the functional dataset, a morphological dataset is created. After rigid alignment of longitudinal and cross-sectional infant T1-w MRI and brain tissue segmentation, we reconstructed and parcellated the cortical surfaces into 35 cortical regions using in-house developed tools \citep{li2014simultaneous}. The connections between these regions contained in the 35$\times$35 morphological graph measure the absolute difference between their cortical thickness. For each subject two morphological graphs are created; one for the left and one for the right hemisphere of the brain. In our experiment, we only focus on one morphological view, hence the final morphological connectivity matrix is created by averaging both hemispheric cortical graphs. We use four different serial time groupings. The first group, denoted as $t_{0}$, includes all brain scans taken between 0-1 months old. The second group, $t_{1}$, contains scans taken between 2-4 months old. The third group, $t_{2}$, contains scans taken between 5-7 months old. Finally, the fourth group, $t_{3}$, contains scans taken between 8-10 months old. Not all subjects have complete availability for all time groups and both modalities. In fact, only 8 subjects possess complete data. Hence, to augment the size of our dataset, we include all subjects that have available data for both modalities at $t_{0}$, resulting in a total of 25 subjects. For subjects with missing data, we apply the imputation techniques proposed during the training process to complete the dataset. However, for the testing dataset, we do not perform any precompletion. Instead, we evaluate only based on the available timepoints and modalities of the real data. 

\begin{figure}[ht!]
\centering
\includegraphics[width=13.5cm]{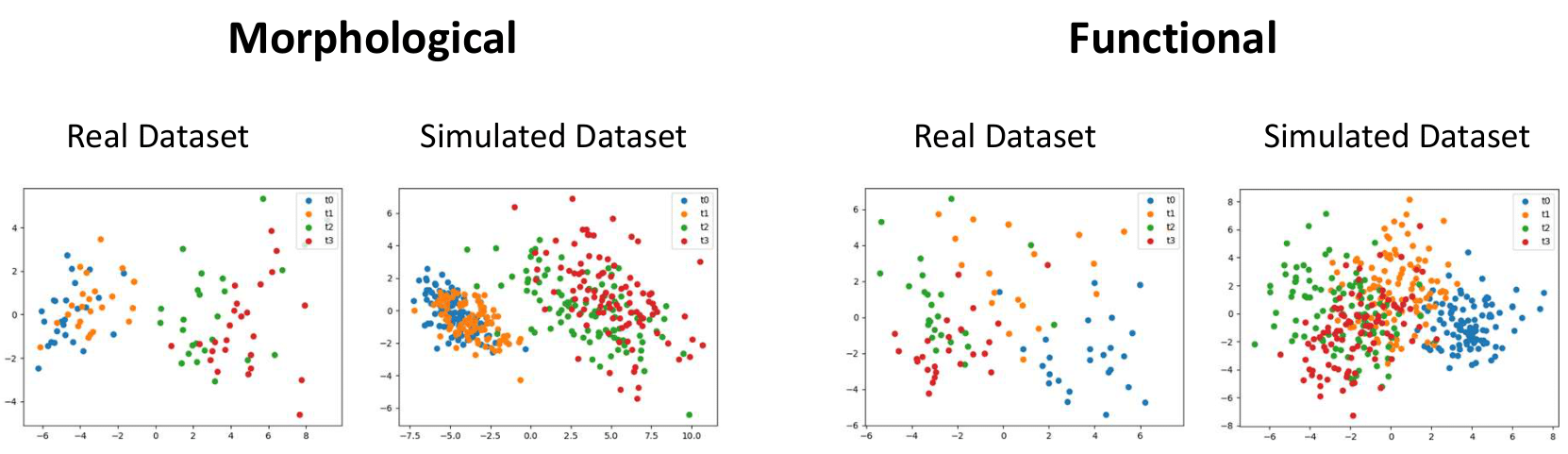}
\caption{PCA comparison between real and simulated datasets. \textbf{Top:} Comparison using the morphological connectome. \textbf{Bottom:} Comparison using the functional connectome.}
\label{fig:pca}
\end{figure}

\begin{figure}[ht!]
\centering
\includegraphics[width=13.5cm]{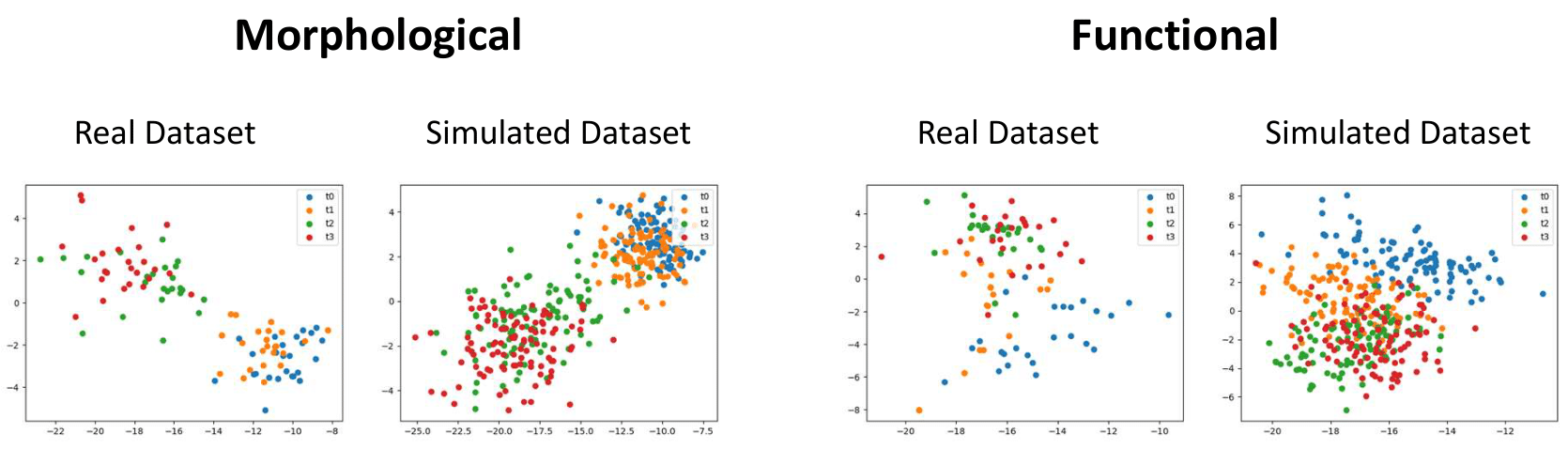}
\caption{SVD comparison between real and simulated datasets. \textbf{Top:} Comparison using the morphological connectome. \textbf{Bottom:} Comparison using the functional connectome.}
\label{fig:svd}
\end{figure}

\textbf{Simulated dataset.} As previously explained, our initial dataset is quite limited in size, representing the data limitation scenario quite well. However, as an additional layer of verification, we also aim to assess how well our framework performs with a slightly larger dataset. To achieve this, we generate a simulated experimental dataset that closely mimics the distribution of the actual data. This simulation involves creating data for 100 distinct subjects by utilizing a multivariate normal distribution, taking into account the mean and covariance values from the original data. Consequently, our 4D simulated dataset is four times the size of the original one. Additionally, since the original dataset contains missing values, we intend to replicate this aspect in our simulated dataset. Given that real-world situations often involve incomplete data, we have chosen to emulate this by randomly omitting 40\% of the values starting from $t_{1}$ onwards. It is important to note that the baseline data at $t_{0}$ remains complete for all subjects since it is a prerequisite for our dataset. By incorporating a 40\% missing rate, we can accurately model and assess a scenario with a substantial amount of absent data. The simulated dataset will be used towards the end of this chapter to achieve a more comprehensive evaluation of the effectiveness of our proposed methods. To confirm that our simulated dataset faithfully mirrors the real dataset, we conduct spectral decomposition on both the real and simulated datasets, allowing for a comprehensive comparison of outcomes. This can help us understand the underlying structure of the original data and examine if the simulated data replicates it. Specifically, we employ two distinct spectral decomposition techniques, Principal Component Analysis (PCA) and Singular Value Decomposition (SVD). The PCA visualizations can be found in \textbf{Fig.} \ref{fig:pca}, while the SVD visualizations are in \textbf{Fig.} \ref{fig:svd}. From the comparison between the real and simulated data in both of the approaches above we can see that the distributions between the original and simulated datasets are very similar.

\subsection{Comparison methods}
In this paper, our goal is to develop a novel brain trajectory evolution network that excels in multi-trajectory prediction. Given our focus on medical data and specifically multi-modality longitudinal brain scans of infants, we want our network to perform effectively when confronted with limited and scarce data.
Our initial proposition involves the integration of the federated learning paradigm. Hence, the primary objective of this paper is to explore whether the adoption of federated learning enhances model quality when compared to the conventional approach of training hospitals solely on their limited datasets. To achieve this, we benchmark our FedGmTE-Net framework, which leverages the FedAvg algorithm, alongside its no-federation counterpart, named NoFedGmTE-Net. Lastly, we aim to conduct a comparative analysis involving our ultimate proposed framework, FedGmTE-Net++, which incorporates specialized mechanisms for addressing data scarcity. We benchmark our model against 1) FedGmTE-Net+, 2) the standard federation framework, FedGmTE-Net, and 3) the non-federation baseline model, NoFedGmTE-Net.

\subsection{Hyperparameter setting and training}
We explore federation between 3 different hospitals and assess the effectiveness of our framework using a 4-fold cross-validation approach. The dataset is divided into four parts. Unless otherwise specified this is a random and uniform split. One fold is serving as the ground-truth test data and the remaining 3 folds are being distributed among the 3 hospitals for local training. To determine the optimal stopping point during training, we employ early stopping with a patience value of 10. This ensures fairness by terminating the training process if there is no significant improvement within the specified limit. The hyperparameter used to monitor the topology loss is set to 0.001. Additionally, for the relevant experiments, the auxiliary regularizer hyperparameter is set to 0.5. 
In the imputation refinement step, we efficiently train the similarity regressors for 2000 epochs before utilizing them to update the imputations. Finally, all hospitals are trained for 5 local epochs before the global aggregation.

In some of the upcoming experiments, we need to test our framework in a heterogeneous scenario. In that case, instead of randomly distributing the training set across hospitals, we use K-means clustering to group the data and assign one cluster to each hospital. This allows us to investigate the performance of our framework in a scenario where there is both statistical and systems heterogeneity across hospitals. Statistical heterogeneity (non-IID case) arises from the differences in data distributions, while systems heterogeneity is created from varying local training dataset sizes in each hospital due to the clusters not being of the same size.

\subsection{Evaluation measures}

For our evaluation, we measure four distinct measures. Firstly, we calculate our primary measure, which is the normal mean average error (MAE) between the predicted graphs and the ground-truth ones. As secondary measures, we use node strength \citep{opsahl2010node}, Pearson correlation coefficient \citep{cohen2009pearson}, and Jaccard distance \citep{levandowsky1971distance}. The MAE among the node strengths (NS) of both graphs helps us evaluate the topological similarity between the predicted brain graphs and the original graphs. The Pearson correlation coefficient (PCC) allows us to quantify the pattern similarity between graphs, where a high positive PCC value indicates a strong similarity in their patterns. Finally, Jaccard distance (JD) is used as a graph distance-based similarity and it describes the proportion of edges that have been removed or added with respect to the total number of edges appearing in either graph. All of the above measures are calculated for each timepoint and modality independently. The equations for node strength, Pearson correlation coefficient, and Jaccard distance for arbitrary time $t$ and hospital $h$ are given below in the respective order.

$$
\text{N}_{r_{i}}(t, h) = \sum_{\substack{r_{j}=1 \\ r_{j} \neq r_{i}}}^{N} \mathbf{A}_{r_{i}r_{j}}^{t,h}
$$
$$
\rho\,(t, h) = \frac{1}{n_{s}^h}\sum_{s=1}^{n_s^h} \frac{\sum_{i=1}^{d} (\mathbf{v}_{t}^{s}[i] - \bar{\mathbf{v}}_{t}^s)(\hat{\mathbf{v}}_{t}^{s}[i] - \bar{\hat{\mathbf{v}}}_{t}^s)}
           {\sqrt{\sum_{i=1}^{d} (\mathbf{v}_{t}^{s}[i] - \bar{\mathbf{v}}_{t}^s)^2} \sqrt{\sum_{i=1}^{d} (\hat{\mathbf{v}}_{t}^{s}[i] - \bar{\hat{\mathbf{v}}}_{t}^s)^2}}
$$
$$
d_{\text {Jaccard }}(t, h)=\frac{1}{n_{s}^h}\sum_{s=1}^{n_s^h} 1-\frac{\sum_{i=1}^{d} \min \left(\mathbf{v}_{t}^{s}[i], \hat{\mathbf{v}}_{t}^{s}[i]\right)}{\sum_{i=1}^{d} \max \left(\mathbf{v}_{t}^{s}[i], \hat{\mathbf{v}}_{t}^{s}[i]\right)}
$$

\subsection{Proof of concept}
We conducted experiments comparing our FedGmTE-Net framework, which employs the FedAvg algorithm and the no federation approach, named NoFedGmTE-Net. The total graph MAE, as well as the NS, PCC, and JD MAEs for each timepoint, hospital, and method, can be found in \textbf{Table \ref{tab:fedVsNo}}. The federated method consistently outperforms the alternative, demonstrating superior performance in all 12/12 scenarios when measured by the primary graph measure. Furthermore, across secondary measures, it is better in 12/12 scenarios for PCC, and 11/12 scenarios for JD, while it seems to be arbitrary for the NS measure with a score of 6/12. Overall, this highlights the significant benefits of federated learning in enhancing model performance.

\begin{table}[ht!]
    \tablefont
    \centering
    \begin{subtable}
        \tablefont
        \centering
        \resizebox{\textwidth}{!}{
        \begin{tabular}{c|c|ccc|ccc}
            & & \multicolumn{3}{c|}{\textcolor{blue}{\textbf{MAE (graph)$\downarrow$}}} & \multicolumn{3}{c}{\textcolor{purple}{\textbf{MAE (NS)$\downarrow$}}} \\
            & \multirow[b]{1}{*}{ Methods } & $h_{1}$ & $h_{2}$ & $h_{3}$
            & $h_{1}$ & $h_{2}$ & $h_{3}$ \\
            \hline 
            
            \multirow{2}{*}{$t_{0}$} & NoFedGmTE-Net & \multicolumn{1}{c|}{0.180 $\pm$ 0.010} & \multicolumn{1}{c|}{0.173 $\pm$ 0.009} & 0.180 $\pm$ 0.016 & \multicolumn{1}{c|}{4.913 $\pm$ 1.479}  & \multicolumn{1}{c|}{4.591 $\pm$ 0.364}  & 4.910 $\pm$ 0.482 \\
            
            & FedGmTE-Net & \multicolumn{1}{c|}{\textcolor{blue}{\textbf{0.169 $\pm$ 0.007}}}& \multicolumn{1}{c|}{\textcolor{blue}{\textbf{0.168 $\pm$ 0.007}}} &
            \textcolor{blue}{\textbf{0.169 $\pm$ 0.007}} & \multicolumn{1}{c|}{\textcolor{purple}{\textbf{4.210 $\pm$ 0.396}}}  & \multicolumn{1}{c|}{\textcolor{purple}{\textbf{4.296 $\pm$ 0.301}}}  & \textcolor{purple}{\textbf{4.259 $\pm$ 0.191}} \\
            \hline 
    
            \multirow{2}{*}{$t_{1}$} & NoFedGmTE-Net & \multicolumn{1}{c|}{0.135 $\pm$ 0.005} & \multicolumn{1}{c|}{0.130 $\pm$ 0.008} & 0.138 $\pm$ 0.003 & \multicolumn{1}{c|}{3.565 $\pm$ 0.546}  & \multicolumn{1}{c|}{\textcolor{purple}{\textbf{3.520 $\pm$ 0.626}}}  & 3.623 $\pm$ 0.620 \\
            
            & FedGmTE-Net & \multicolumn{1}{c|}{\textcolor{blue}{\textbf{0.125 $\pm$ 0.008}}} & \multicolumn{1}{c|}{\textcolor{blue}{\textbf{0.123 $\pm$ 0.008}}} &
            \textcolor{blue}{\textbf{0.126 $\pm$ 0.008}} & \multicolumn{1}{c|}{\textcolor{purple}{\textbf{3.421 $\pm$ 0.614}}}  & \multicolumn{1}{c|}{3.534 $\pm$ 0.767}  & \textcolor{purple}{\textbf{3.606 $\pm$ 1.008}} \\
            \hline
            
            \multirow{2}{*}{$t_{2}$} & NoFedGmTE-Net & \multicolumn{1}{c|}{0.169 $\pm$ 0.009} & \multicolumn{1}{c|}{0.165 $\pm$ 0.015} & 0.179 $\pm$ 0.010 & \multicolumn{1}{c|}{4.596 $\pm$ 0.798}  & \multicolumn{1}{c|}{\textcolor{purple}{\textbf{4.606 $\pm$ 0.516}}}  & \textcolor{purple}{\textbf{4.815 $\pm$ 0.911}} \\
            
            & FedGmTE-Net & \multicolumn{1}{c|}{\textcolor{blue}{\textbf{0.160 $\pm$ 0.013}}} & \multicolumn{1}{c|}{\textcolor{blue}{\textbf{0.162 $\pm$ 0.015}}} &
            \textcolor{blue}{\textbf{0.166 $\pm$ 0.018}} & \multicolumn{1}{c|}{\textcolor{purple}{\textbf{4.370 $\pm$ 1.134}}}  & \multicolumn{1}{c|}{5.087 $\pm$ 0.917}  & 5.009 $\pm$ 1.698 \\
            \hline
    
            \multirow{2}{*}{$t_{3}$} & NoFedGmTE-Net & \multicolumn{1}{c|}{0.171 $\pm$ 0.008} & \multicolumn{1}{c|}{0.160 $\pm$ 0.006} & 0.175 $\pm$ 0.015 & \multicolumn{1}{c|}{\textcolor{purple}{\textbf{3.824 $\pm$ 0.851}}}  & \multicolumn{1}{c|}{\textcolor{purple}{\textbf{3.427 $\pm$ 0.353}}}  & \textcolor{purple}{\textbf{3.857 $\pm$ 0.547}} \\
            
            & FedGmTE-Net & \multicolumn{1}{c|}{\textcolor{blue}{\textbf{0.163 $\pm$ 0.015}}} & \multicolumn{1}{c|}{\textcolor{blue}{\textbf{0.155 $\pm$ 0.008}}} &
            \textcolor{blue}{\textbf{0.157 $\pm$ 0.011}} & \multicolumn{1}{c|}{4.059 $\pm$ 0.804}  & \multicolumn{1}{c|}{4.150 $\pm$ 0.863}  & 4.216 $\pm$ 0.543 \\
            \hline
        \end{tabular}
        }

    \end{subtable}
    
    \begin{subtable}
        \tablefont
        \centering
        \resizebox{\textwidth}{!}{
        \begin{tabular}{c|c|ccc|ccc}
            & & \multicolumn{3}{c|}{\textcolor{gold}{\textbf{MAE (PCC)$\uparrow$}}} & \multicolumn{3}{c}{\textcolor{green}{\textbf{MAE (JD)$\downarrow$}}} \\
            & \multirow[b]{1}{*}{ Methods } & $h_{1}$ & $h_{2}$ & $h_{3}$
            & $h_{1}$ & $h_{2}$ & $h_{3}$ \\
            \hline 
            
            \multirow{2}{*}{$t_{0}$} & NoFedGmTE-Net & \multicolumn{1}{c|}{0.260 $\pm$ 0.076} & \multicolumn{1}{c|}{0.330 $\pm$ 0.061} & 0.253 $\pm$ 0.111 & \multicolumn{1}{c|}{0.647 $\pm$ 0.038}  & \multicolumn{1}{c|}{0.627 $\pm$ 0.017}  & 0.649 $\pm$ 0.026 \\
            
            & FedGmTE-Net & \multicolumn{1}{c|}{\textcolor{gold}{\textbf{0.380 $\pm$ 0.020}}} & \multicolumn{1}{c|}{\textcolor{gold}{\textbf{0.385 $\pm$ 0.018}}} &
            \textcolor{gold}{\textbf{0.382 $\pm$ 0.020}} & \multicolumn{1}{c|}{\textcolor{green}{\textbf{0.611 $\pm$ 0.007}}}  & \multicolumn{1}{c|}{\textcolor{green}{\textbf{0.612 $\pm$ 0.006}}}  & \textcolor{green}{\textbf{0.612 $\pm$ 0.010}} \\
            \hline 
    
            \multirow{2}{*}{$t_{1}$} & NoFedGmTE-Net & \multicolumn{1}{c|}{0.680$\pm$ 0.012} & \multicolumn{1}{c|}{0.681 $\pm$ 0.015} & 0.655 $\pm$ 0.022 & \multicolumn{1}{c|}{0.477 $\pm$ 0.018}  & \multicolumn{1}{c|}{0.477 $\pm$ 0.015}  & 0.488 $\pm$ 0.012 \\
            
            & FedGmTE-Net & \multicolumn{1}{c|}{\textcolor{gold}{\textbf{0.717 $\pm$ 0.007}}} & \multicolumn{1}{c|}{\textcolor{gold}{\textbf{0.713 $\pm$ 0.012}}} &
            \textcolor{gold}{\textbf{0.711 $\pm$ 0.004}} & \multicolumn{1}{c|}{\textcolor{green}{\textbf{0.449 $\pm$ 0.018}}}  & \multicolumn{1}{c|}{\textcolor{green}{\textbf{0.457 $\pm$ 0.019}}}  & \textcolor{green}{\textbf{0.458 $\pm$ 0.024}} \\
            \hline
            
            \multirow{2}{*}{$t_{2}$} & NoFedGmTE-Net & \multicolumn{1}{c|}{0.688 $\pm$ 0.023} & \multicolumn{1}{c|}{0.695 $\pm$ 0.017} & 0.673 $\pm$ 0.006 & \multicolumn{1}{c|}{0.474 $\pm$ 0.021}  & \multicolumn{1}{c|}{0.470 $\pm$ 0.016}  & 0.486 $\pm$ 0.014 \\
            
            & FedGmTE-Net & \multicolumn{1}{c|}{\textcolor{gold}{\textbf{0.720 $\pm$ 0.018}}} & \multicolumn{1}{c|}{\textcolor{gold}{\textbf{0.715 $\pm$ 0.020}}} &
            \textcolor{gold}{\textbf{0.712 $\pm$ 0.019}} & \multicolumn{1}{c|}{\textcolor{green}{\textbf{0.449 $\pm$ 0.028}}}  & \multicolumn{1}{c|}{\textcolor{green}{\textbf{0.467 $\pm$ 0.021}}}  & \textcolor{green}{\textbf{0.468 $\pm$ 0.038}} \\
            \hline
    
            \multirow{2}{*}{$t_{3}$} & NoFedGmTE-Net & \multicolumn{1}{c|}{0.691 $\pm$ 0.009} & \multicolumn{1}{c|}{0.701 $\pm$ 0.016} & 0.663 $\pm$ 0.033 & \multicolumn{1}{c|}{0.463 $\pm$ 0.009}  & \multicolumn{1}{c|}{\textcolor{green}{\textbf{0.451 $\pm$ 0.009}}}  & 0.649 $\pm$ 0.024 \\
            
            & FedGmTE-Net & \multicolumn{1}{c|}{\textcolor{gold}{\textbf{0.726 $\pm$ 0.020}}} & \multicolumn{1}{c|}{\textcolor{gold}{\textbf{0.722 $\pm$ 0.021}}} &
            \textcolor{gold}{\textbf{0.714 $\pm$ 0.039}} & \multicolumn{1}{c|}{\textcolor{green}{\textbf{0.442 $\pm$ 0.013}}}  & \multicolumn{1}{c|}{0.452 $\pm$ 0.023}  & \textcolor{green}{\textbf{0.628 $\pm$ 0.030}} \\
            \hline
        \end{tabular}
        }
    \end{subtable}
    \caption{Comparison of FedGmTE-Net and its no federation counterpart
    (NoFedGmTE-Net). We highlight in bold, blue for MAE (graph), purple for MAE (NS), gold for MAE (PCC), and green for MAE (JD) the best performance at each hospital across timepoints. The performances for the morphological and functional trajectories are averaged.}
    \label{tab:fedVsNo}
\end{table}

\subsection{Performance evaluation}

\begin{figure}[htb!]
\centering
{\includegraphics[width=13.5cm]{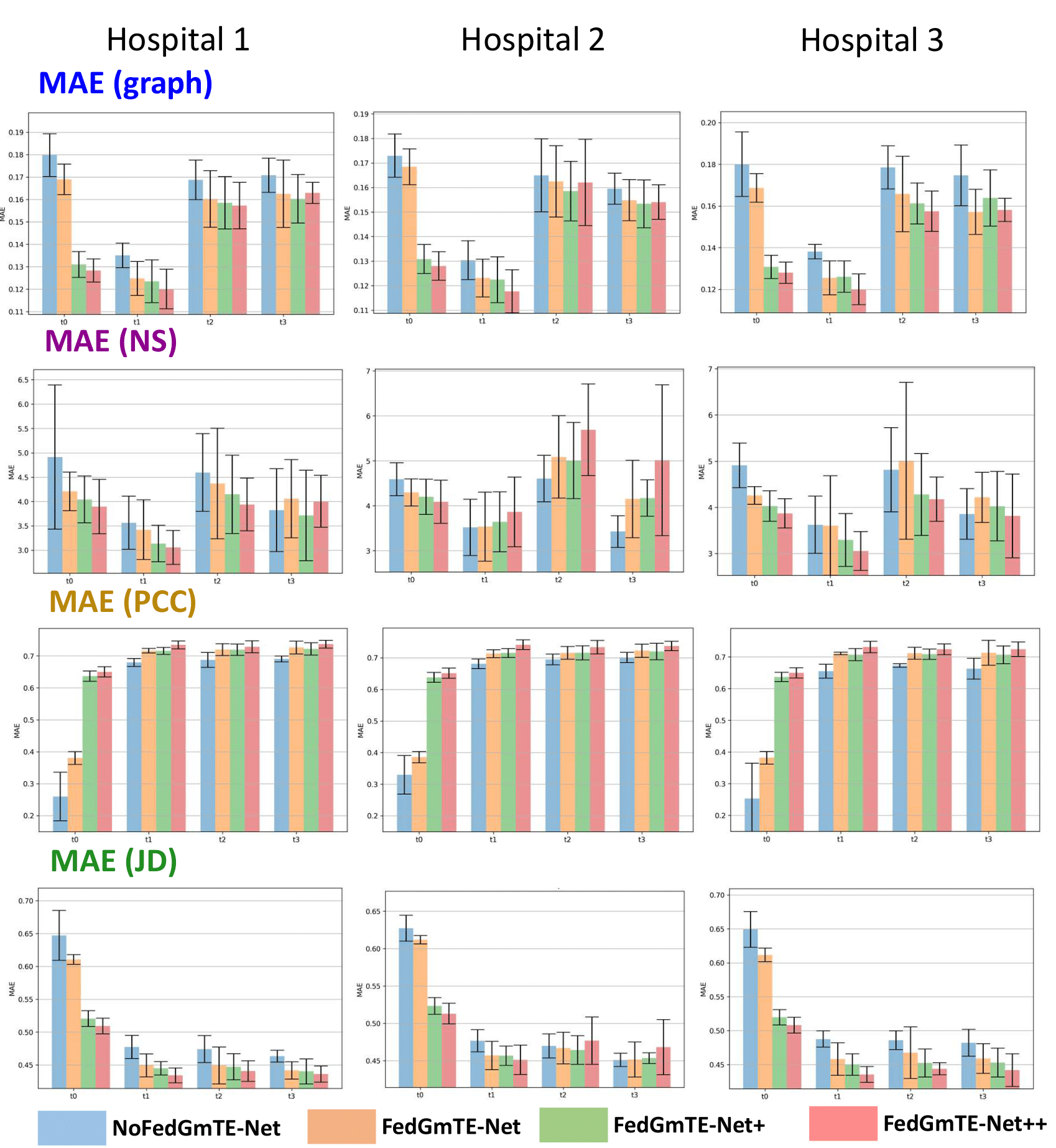}}
\caption{\textbf{Real IID data}: Total MAE (graph), MAE (NS), MAE (PCC), and MAE (JD) recorded for the three hospitals at all timepoints. The average value across different modalities was taken for the NoFedGmTE-Net, FedGmTE-Net, FedGmTE-Net+, and FedGmTE-Net++ frameworks.}
\label{fig:real_iid_graph}
\end{figure}

We further assess our upgraded proposed FedGmTE-Net++, against its earlier version FedGmTE-Net+, the conventional federation framework, FedGmTE-Net, and the non-federation baseline model, NoFedGmTE-Net. The total MAE plots for all the measures across modalities are provided in \textbf{Fig.} \ref{fig:real_iid_graph}. These plots clearly demonstrate the consistent superiority of our proposed FedGmTE-Net++ over other methods across various timepoints and modalities, with FedGmTE-Net+ emerging as the second-best performer in most instances. Specifically, when considering the MAE (graph) measure, FedGmTE-Net++ outperforms the alternatives in 9/12 scenarios. In terms of the PCC measure, FedGmTE-Net++ consistently demonstrates superior performance in all 12/12 scenarios, while for MAE (JD), it excels in 10/12 scenarios. The NS measure seems to yield slightly worse results, yet FedGmTE-Net++ still stands out as the better performer in 8/12 scenarios. 

\begin{figure}[htb!]
\centering
{\includegraphics[width=13.5cm]{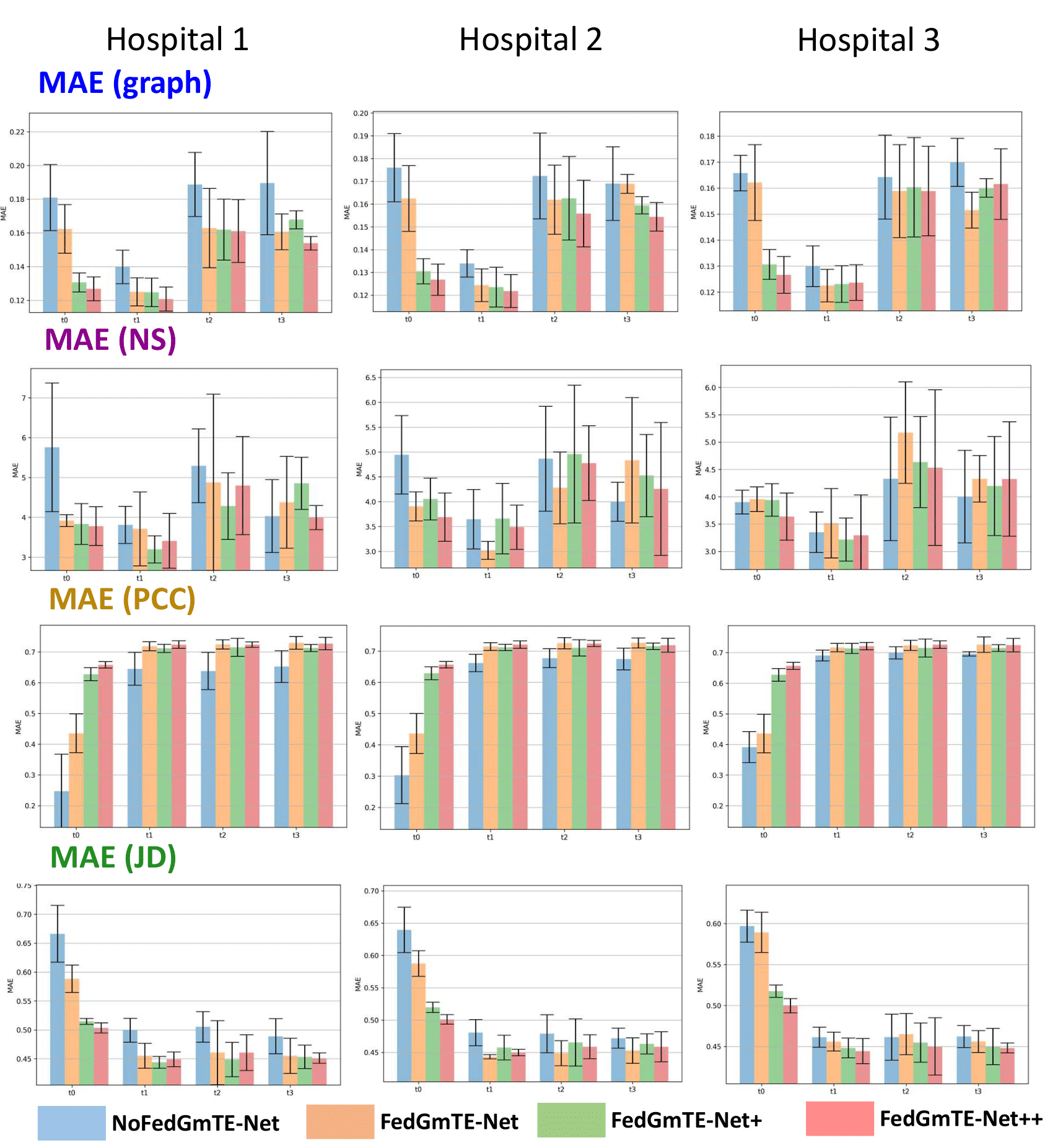}}
\caption{\textbf{Real Non-IID data}: Total MAE (graph), MAE (NS), MAE (PCC), and MAE (JD) recorded for the three hospitals at all timepoints. The average value across different modalities was taken for the NoFedGmTE-Net, FedGmTE-Net, FedGmTE-Net+, and FedGmTE-Net++ frameworks.}
\label{fig:real_non_iid_graph}
\end{figure}

In the next experiment, instead of randomly distributing the training set across hospitals, we use K-means clustering to group the data and assign one cluster to each hospital. This allows us to investigate the performance of our framework in a scenario where there is both statistical (non-IID case) and systems heterogeneity across hospitals. The total MAE plots averaged across modalities are provided in \textbf{Fig.} \ref{fig:real_non_iid_graph}. Our findings are similar to the IID case, with a slight degradation. In terms of MAE, FedGmTE-Net++ maintains its consistent superiority over the alternatives in 10/12 scenarios. However, for the secondary measures, the performance shows a slight dip, with FedGmTE-Net++ still outperforming the others in the majority of scenarios; 7/12 for JD and 8/12 for PCC, while the NS measure yields very inconsistent results across methods (4/12). Despite this minor dip in results compared to the IID case, the above shows that our proposed framework (FedGmTE-Net++) outperforms the benchmarks even in a heterogeneous setting, highlighting the robustness of our method. For a comprehensive and in-depth examination of the outcomes for each experiment conducted separately within each modality, refer to the following tables: \textbf{Table \ref{tab:real_iid_morph}} and \textbf{Table \ref{tab:real_iid_func}} for the IID experiments, and \textbf{Table \ref{tab:real_non_iid_morph}} and \textbf{Table \ref{tab:real_non_iid_func}} for the non-IID experiments. These tables break down the results, providing a perspective on the performance of our framework in each specific modality and timepoint individually.

\begin{table}[ht!]
    \tablefont
    \centering
    \begin{subtable}
        \tablefont
        \centering
        \resizebox{\textwidth}{!}{
\begin{tabular}{c|c|ccc|ccc}
            & & \multicolumn{3}{c|}{\textcolor{blue}{\textbf{MAE (graph)$\downarrow$}}} & \multicolumn{3}{c}{\textcolor{purple}{\textbf{MAE (NS)$\downarrow$}}} \\
            & \multirow[b]{1}{*}{ Methods } & $h_{1}$ & $h_{2}$ & $h_{3}$
            & $h_{1}$ & $h_{2}$ & $h_{3}$ \\
            \hline 
            
            \multirow{4}{*}{$t_{0}$} & NoFedGmTE-Net & \multicolumn{1}{c|}{0.215 $\pm$ 0.014} & \multicolumn{1}{c|}{0.203 $\pm$ 0.017} & 0.216 $\pm$ 0.025 & \multicolumn{1}{c|}{4.479 $\pm$ 0.742}  & \multicolumn{1}{c|}{4.259 $\pm$ 0.493}  & 4.427 $\pm$ 0.667 \\
            
            & FedGmTE-Net & \multicolumn{1}{c|}{0.196 $\pm$ 0.012} & \multicolumn{1}{c|}{0.195 $\pm$ 0.014} &
            0.196 $\pm$ 0.013 & \multicolumn{1}{c|}{3.834 $\pm$ 0.241}  & \multicolumn{1}{c|}{3.953 $\pm$ 0.354}  & 3.906 $\pm$ 0.412 \\
            
            & FedGmTE-Net+ & \multicolumn{1}{c|}{0.140 $\pm$ 0.012} & \multicolumn{1}{c|}{0.140 $\pm$ 0.012} & 0.139 $\pm$ 0.011 & \multicolumn{1}{c|}{2.290 $\pm$ 0.232} & \multicolumn{1}{c|}{2.374 $\pm$ 0.340} & 2.275 $ \pm$ 0.313  \\ 
    
            & FedGmTE-Net++ & \multicolumn{1}{c|}{\textcolor{blue}{\textbf{0.137 $\pm$ 0.011}}} & \multicolumn{1}{c|}{\textcolor{blue}{\textbf{0.137 $\pm$ 0.012}}} & \textcolor{blue}{\textbf{0.136 $\pm$ 0.011}} & \multicolumn{1}{c|}{\textcolor{purple}{\textbf{2.273 $\pm$ 0.090}}} & \multicolumn{1}{c|}{\textcolor{purple}{\textbf{2.366 $\pm$ 0.292}}} & \textcolor{purple}{\textbf{2.230 $ \pm$ 0.289}}  \\
            \hline 
    
            \multirow{4}{*}{$t_{1}$} & NoFedGmTE-Net & \multicolumn{1}{c|}{0.147 $\pm$ 0.009} & \multicolumn{1}{c|}{0.142 $\pm$ 0.016} & 0.150 $\pm$ 0.008 & \multicolumn{1}{c|}{2.465 $\pm$ 0.420}  & \multicolumn{1}{c|}{2.420 $\pm$ 0.350}  & 2.426 $\pm$ 0.333 \\
\hline
        \end{tabular}
        }

    \end{subtable}
    
    \begin{subtable}
        \tablefont
        \centering
         \resizebox{\textwidth}{!}{
        \begin{tabular}{c|c|ccc|ccc}
            & & \multicolumn{3}{c|}{\textcolor{gold}{\textbf{MAE (PC)$\uparrow$}}} & \multicolumn{3}{c}{\textcolor{green}{\textbf{MAE (JD)$\downarrow$}}} \\
            & \multirow[b]{1}{*}{ Methods } & $h_{1}$ & $h_{2}$ & $h_{3}$
            & $h_{1}$ & $h_{2}$ & $h_{3}$ \\
            \hline 
            
            \multirow{4}{*}{$t_{0}$} & NoFedGmTE-Net & \multicolumn{1}{c|}{0.310 $\pm$ 0.119} & \multicolumn{1}{c|}{0.411 $\pm$ 0.091} & 0.319 $\pm$ 0.152 & \multicolumn{1}{c|}{0.542 $\pm$ 0.331}  & \multicolumn{1}{c|}{0.514 $\pm$ 0.029}  & 0.541 $\pm$ 0.038 \\
            
            & FedGmTE-Net & \multicolumn{1}{c|}{0.475 $\pm$ 0.028} & \multicolumn{1}{c|}{0.483 $\pm$ 0.021} &
            0.479 $\pm$ 0.025 & \multicolumn{1}{c|}{0.492 $\pm$ 0.014}  & \multicolumn{1}{c|}{0.494 $\pm$ 0.014}  & 0.493 $\pm$ 0.018 \\
            
            & FedGmTE-Net+ & \multicolumn{1}{c|}{0.763 $\pm$ 0.011} & \multicolumn{1}{c|}{0.765 $\pm$ 0.013} & 0.765 $\pm$ 0.012 & \multicolumn{1}{c|}{0.349 $\pm$ 0.020} & \multicolumn{1}{c|}{0.351 $\pm$ 0.020} & 0.347 $\pm$ 0.017  \\ 
    
            & FedGmTE-Net++ & \multicolumn{1}{c|}{\textcolor{gold}{\textbf{0.769 $\pm$ 0.017}}} & \multicolumn{1}{c|}{\textcolor{gold}{\textbf{0.772 $\pm$ 0.018}}} & \textcolor{gold}{\textbf{0.770 $\pm$ 0.019}} & \multicolumn{1}{c|}{\textcolor{green}{\textbf{0.342 $\pm$ 0.016}}} & \multicolumn{1}{c|}{\textcolor{green}{\textbf{0.344 $\pm$ 0.022}}} & \textcolor{green}{\textbf{0.340 $ \pm$ 0.018}}  \\
            \hline 
    
            \multirow{4}{*}{$t_{1}$} & NoFedGmTE-Net & \multicolumn{1}{c|}{0.796$\pm$ 0.010} & \multicolumn{1}{c|}{0.799 $\pm$ 0.038} & 0.774 $\pm$ 0.002 & \multicolumn{1}{c|}{0.331 $\pm$ 0.022}  & \multicolumn{1}{c|}{0.323 $\pm$ 0.024}  & 0.336 $\pm$ 0.008 \\
\hline
        \end{tabular}
        }
    \end{subtable}
    \caption{\textbf{Real IID data}: Morphological brain graph trajectory prediction by NoFedGmTE-Net, FedGmTE-Net, FedGmTE-Net+ and FedGmTE-Net++. We highlight in bold, blue for MAE (graph), purple for MAE (NS), gold for MAE (PC), and green for MAE (JD) the best performance at each hospital across timepoints.}
    \label{tab:real_iid_morph}
\end{table}

\begin{table}
    \tablefont
    \centering
    \begin{subtable}
        \tablefont
        \centering
         \resizebox{\textwidth}{!}{
        \begin{tabular}{c|c|ccc|ccc}
            & & \multicolumn{3}{c|}{\textcolor{blue}{\textbf{MAE (graph)$\downarrow$}}} & \multicolumn{3}{c}{\textcolor{purple}{\textbf{MAE (NS)$\downarrow$}}} \\
            & \multirow[b]{1}{*}{ Methods } & $h_{1}$ & $h_{2}$ & $h_{3}$
            & $h_{1}$ & $h_{2}$ & $h_{3}$ \\
            \hline 
            
            \multirow{4}{*}{$t_{0}$} & NoFedGmTE-Net & \multicolumn{1}{c|}{0.144 $\pm$ 0.005} & \multicolumn{1}{c|}{0.143 $\pm$ 0.005} & 0.144 $\pm$ 0.008 & \multicolumn{1}{c|}{5.346 $\pm$ 2.289}  & \multicolumn{1}{c|}{4.924 $\pm$ 0.876}  & 5.393 $\pm$ 0.849 \\
            
            & FedGmTE-Net & \multicolumn{1}{c|}{0.142 $\pm$ 0.005} & \multicolumn{1}{c|}{0.142 $\pm$ 0.005} &
            0.142 $\pm$ 0.005 & \multicolumn{1}{c|}{\textcolor{purple}{\textbf{4.586 $\pm$ 0.954}}}  & \multicolumn{1}{c|}{\textcolor{purple}{\textbf{4.639 $\pm$ 0.800}}}  & \textcolor{purple}{\textbf{4.612 $\pm$ 0.794}} \\
            
            & FedGmTE-Net+ & \multicolumn{1}{c|}{0.122 $\pm$ 0.003} & \multicolumn{1}{c|}{0.122 $\pm$ 0.003} & 0.122 $\pm$ 0.003 & \multicolumn{1}{c|}{5.796 $\pm$ 1.056} & \multicolumn{1}{c|}{6.025 $\pm$ 0.897} & 5.779 $ \pm$ 0.911  \\ 
    
            & FedGmTE-Net++ & \multicolumn{1}{c|}{\textcolor{blue}{\textbf{0.120 $\pm$ 0.004}}} & \multicolumn{1}{c|}{\textcolor{blue}{\textbf{0.119 $\pm$ 0.004}}} & \textcolor{blue}{\textbf{0.120 $\pm$ 0.004}} & \multicolumn{1}{c|}{5.519 $\pm$ 1.172} & \multicolumn{1}{c|}{5.812 $\pm$ 1.034} & 5.507 $ \pm$ 0.879  \\
            \hline 
    
            \multirow{4}{*}{$t_{1}$} & NoFedGmTE-Net & \multicolumn{1}{c|}{0.123 $\pm$ 0.004} & \multicolumn{1}{c|}{0.119 $\pm$ 0.003} & 0.127 $\pm$ 0.007 & \multicolumn{1}{c|}{4.666 $\pm$ 0.678}  & \multicolumn{1}{c|}{\textcolor{purple}{\textbf{4.612 $\pm$ 1.141}}}  & 4.820 $\pm$ 1.153 \\
\hline
        \end{tabular}
        }

    \end{subtable}
    
    \begin{subtable}
        \tablefont
        \centering
         \resizebox{\textwidth}{!}{
        \begin{tabular}{c|c|ccc|ccc}
            & & \multicolumn{3}{c|}{\textcolor{gold}{\textbf{MAE (PC)$\uparrow$}}} & \multicolumn{3}{c}{\textcolor{green}{\textbf{MAE (JD)$\downarrow$}}} \\
            & \multirow[b]{1}{*}{ Methods } & $h_{1}$ & $h_{2}$ & $h_{3}$
            & $h_{1}$ & $h_{2}$ & $h_{3}$ \\
            \hline 
            
            \multirow{4}{*}{$t_{0}$} & NoFedGmTE-Net & \multicolumn{1}{c|}{0.210 $\pm$ 0.034} & \multicolumn{1}{c|}{0.248 $\pm$ 0.037} & 0.187 $\pm$ 0.074 & \multicolumn{1}{c|}{0.752 $\pm$ 0.035}  & \multicolumn{1}{c|}{0.740 $\pm$ 0.012}  & 0.758 $\pm$ 0.017 \\
            
            & FedGmTE-Net & \multicolumn{1}{c|}{0.285 $\pm$ 0.027} & \multicolumn{1}{c|}{0.288 $\pm$ 0.029} &
            0.285 $\pm$ 0.028 & \multicolumn{1}{c|}{0.729 $\pm$ 0.009}  & \multicolumn{1}{c|}{0.730 $\pm$ 0.005}  & 0.730 $\pm$ 0.008 \\
            
            & FedGmTE-Net+ & \multicolumn{1}{c|}{0.510 $\pm$ 0.027} & \multicolumn{1}{c|}{0.511 $\pm$ 0.026} & 0.509 $\pm$ 0.025 & \multicolumn{1}{c|}{0.692 $\pm$ 0.014} & \multicolumn{1}{c|}{0.696 $\pm$ 0.011} & 0.692 $\pm$ 0.015  \\ 
    
            & FedGmTE-Net++ & \multicolumn{1}{c|}{\textcolor{gold}{\textbf{0.530 $\pm$ 0.020}}} & \multicolumn{1}{c|}{\textcolor{gold}{\textbf{0.530 $\pm$ 0.019}}} & \textcolor{gold}{\textbf{0.529 $\pm$ 0.019}} & \multicolumn{1}{c|}{\textcolor{green}{\textbf{0.676 $\pm$ 0.015}}} & \multicolumn{1}{c|}{\textcolor{green}{\textbf{0.682 $\pm$ 0.012}}} & \textcolor{green}{\textbf{0.677 $ \pm$ 0.012}}  \\
            \hline 
    
            \multirow{4}{*}{$t_{1}$} & NoFedGmTE-Net & \multicolumn{1}{c|}{0.563 $\pm$ 0.019} & \multicolumn{1}{c|}{0.563 $\pm$ 0.022} & 0.536 $\pm$ 0.044 & \multicolumn{1}{c|}{0.624 $\pm$ 0.016}  & \multicolumn{1}{c|}{0.630 $\pm$ 0.009}  & 0.640 $\pm$ 0.022 \\
\hline
        \end{tabular}
        }
    \end{subtable}
    \caption{\textbf{Real IID data}: Functional brain graph trajectory prediction by NoFedGmTE-Net, FedGmTE-Net, FedGmTE-Net+ and FedGmTE-Net++. We highlight in bold, blue for MAE (graph), purple for MAE (NS), gold for MAE (PC), and green for MAE (JD) the best performance at each hospital across timepoints.}
    \label{tab:real_iid_func}
\end{table}

\begin{table}
    \tablefont
    \centering
    \begin{subtable}
        \tablefont
        \centering
        \resizebox{\textwidth}{!}{
        \begin{tabular}{c|c|ccc|ccc}
            & & \multicolumn{3}{c|}{\textcolor{blue}{\textbf{MAE (graph)$\downarrow$}}} & \multicolumn{3}{c}{\textcolor{purple}{\textbf{MAE (NS)$\downarrow$}}} \\
            & \multirow[b]{1}{*}{ Methods } & $h_{1}$ & $h_{2}$ & $h_{3}$
            & $h_{1}$ & $h_{2}$ & $h_{3}$ \\
            \hline 
            
            \multirow{4}{*}{$t_{0}$} & NoFedGmTE-Net & \multicolumn{1}{c|}{0.219 $\pm$ 0.035} & \multicolumn{1}{c|}{0.209 $\pm$ 0.028} & 0.186 $\pm$ 0.012 & \multicolumn{1}{c|}{4.935 $\pm$ 1.256}  & \multicolumn{1}{c|}{4.264 $\pm$ 0.997}  & 3.522 $\pm$ 0.361 \\
            
            & FedGmTE-Net & \multicolumn{1}{c|}{0.184 $\pm$ 0.026} & \multicolumn{1}{c|}{0.184 $\pm$ 0.026} &
            0.184 $\pm$ 0.026 & \multicolumn{1}{c|}{3.502 $\pm$ 0.707}  & \multicolumn{1}{c|}{3.486 $\pm$ 0.639}  & 3.530 $\pm$ 0.786 \\
            
            & FedGmTE-Net+ & \multicolumn{1}{c|}{0.138 $\pm$ 0.011} & \multicolumn{1}{c|}{0.139 $\pm$ 0.011} & 0.139 $\pm$ 0.012 & \multicolumn{1}{c|}{\textcolor{purple}{\textbf{2.288 $\pm$ 0.208}}} & \multicolumn{1}{c|}{2.398 $\pm$ 0.243} & 2.353 $ \pm$ 0.331  \\ 
    
            & FedGmTE-Net++ & \multicolumn{1}{c|}{\textcolor{blue}{\textbf{0.133 $\pm$ 0.014}}} & \multicolumn{1}{c|}{\textcolor{blue}{\textbf{0.133 $\pm$ 0.014}}} & \textcolor{blue}{\textbf{0.133 $\pm$ 0.014}} & \multicolumn{1}{c|}{2.327 $\pm$ 0.278} & \multicolumn{1}{c|}{\textcolor{purple}{\textbf{2.218 $\pm$ 0.284}}} & \textcolor{purple}{\textbf{2.217 $ \pm$ 0.206}}  \\
            \hline 
    
            \multirow{4}{*}{$t_{1}$} & NoFedGmTE-Net & \multicolumn{1}{c|}{0.159 $\pm$ 0.018} & \multicolumn{1}{c|}{0.144 $\pm$ 0.015} & 0.140 $\pm$ 0.014 & \multicolumn{1}{c|}{2.661 $\pm$ 0.233}  & \multicolumn{1}{c|}{2.326 $\pm$ 0.464}  & 2.254 $\pm$ 0.385 \\
\hline
        \end{tabular}
        }

    \end{subtable}
    
    \begin{subtable}
        \tablefont
        \centering
        \resizebox{\textwidth}{!}{
        \begin{tabular}{c|c|ccc|ccc}
            & & \multicolumn{3}{c|}{\textcolor{gold}{\textbf{MAE (PC)$\uparrow$}}} & \multicolumn{3}{c}{\textcolor{green}{\textbf{MAE (JD)$\downarrow$}}} \\
            & \multirow[b]{1}{*}{ Methods } & $h_{1}$ & $h_{2}$ & $h_{3}$
            & $h_{1}$ & $h_{2}$ & $h_{3}$ \\
            \hline 
            
            \multirow{4}{*}{$t_{0}$} & NoFedGmTE-Net & \multicolumn{1}{c|}{0.307 $\pm$ 0.178} & \multicolumn{1}{c|}{0.397 $\pm$ 0.136} & 0.532 $\pm$ 0.061 & \multicolumn{1}{c|}{0.556 $\pm$ 0.065}  & \multicolumn{1}{c|}{0.523 $\pm$ 0.055}  & 0.468 $\pm$ 0.026 \\
            
            & FedGmTE-Net & \multicolumn{1}{c|}{0.547 $\pm$ 0.090} & \multicolumn{1}{c|}{0.548 $\pm$ 0.090} &
            0.547 $\pm$ 0.090 & \multicolumn{1}{c|}{0.461 $\pm$ 0.043}  & \multicolumn{1}{c|}{0.460 $\pm$ 0.040}  & 0.462 $\pm$ 0.045 \\
            
            & FedGmTE-Net+ & \multicolumn{1}{c|}{0.745 $\pm$ 0.029} & \multicolumn{1}{c|}{0.747 $\pm$ 0.029} & 0.745 $\pm$ 0.030 & \multicolumn{1}{c|}{0.344 $\pm$ 0.015} & \multicolumn{1}{c|}{0.348 $\pm$ 0.016} & 0.347 $ \pm$ 0.018  \\ 
    
            & FedGmTE-Net++ & \multicolumn{1}{c|}{\textcolor{gold}{\textbf{0.791 $\pm$ 0.021}}} & \multicolumn{1}{c|}{\textcolor{gold}{\textbf{0.787 $\pm$ 0.021}}} & \textcolor{gold}{\textbf{0.789 $\pm$ 0.022}} & \multicolumn{1}{c|}{\textcolor{green}{\textbf{0.335 $\pm$ 0.020}}} & \multicolumn{1}{c|}{\textcolor{green}{\textbf{0.332 $\pm$ 0.019}}} & \textcolor{green}{\textbf{0.331 $ \pm$ 0.018}}  \\
            \hline 
    
            \multirow{4}{*}{$t_{1}$} & NoFedGmTE-Net & \multicolumn{1}{c|}{0.735 $\pm$ 0.063} & \multicolumn{1}{c|}{0.782 $\pm$ 0.049} & 0.797 $\pm$ 0.028 & \multicolumn{1}{c|}{0.359 $\pm$ 0.027}  & \multicolumn{1}{c|}{0.326 $\pm$ 0.036}  & 0.312 $\pm$ 0.024 \\
\hline
        \end{tabular}
        }
    \end{subtable}
    \caption{\textbf{Real Non-IID data}: Morphological brain graph trajectory prediction by NoFedGmTE-Net, FedGmTE-Net, FedGmTE-Net+ and FedGmTE-Net++. We highlight in bold, blue for MAE (graph), purple for MAE (NS), gold for MAE (PC), and green for MAE (JD) the best performance at each hospital across timepoints.}
    \label{tab:real_non_iid_morph}
\end{table}

\begin{table}
    \tablefont
    \centering
    \begin{subtable}
        \tablefont
        \centering
        \resizebox{\textwidth}{!}{
        \begin{tabular}{c|c|ccc|ccc}
            & & \multicolumn{3}{c|}{\textcolor{blue}{\textbf{MAE (graph)$\downarrow$}}} & \multicolumn{3}{c}{\textcolor{purple}{\textbf{MAE (NS)$\downarrow$}}} \\
            & \multirow[b]{1}{*}{ Methods } & $h_{1}$ & $h_{2}$ & $h_{3}$
            & $h_{1}$ & $h_{2}$ & $h_{3}$ \\
            \hline 
            
            \multirow{4}{*}{$t_{0}$} & NoFedGmTE-Net & \multicolumn{1}{c|}{0.143 $\pm$ 0.006} & \multicolumn{1}{c|}{0.143 $\pm$ 0.004} & 0.145 $\pm$ 0.005 & \multicolumn{1}{c|}{6.577 $\pm$ 2.345}  & \multicolumn{1}{c|}{5.623 $\pm$ 0.955}  & \textcolor{purple}{\textbf{4.284 $\pm$ 0.682}} \\
            
            & FedGmTE-Net & \multicolumn{1}{c|}{0.141 $\pm$ 0.005} & \multicolumn{1}{c|}{0.141 $\pm$ 0.005} &
            0.140 $\pm$ 0.005 & \multicolumn{1}{c|}{\textcolor{purple}{\textbf{4.336 $\pm$ 0.602}}}  & \multicolumn{1}{c|}{\textcolor{purple}{\textbf{4.323 $\pm$ 0.758}}}  & 4.379 $\pm$ 0.602 \\
            
            & FedGmTE-Net+ & \multicolumn{1}{c|}{0.123 $\pm$ 0.003} & \multicolumn{1}{c|}{0.122 $\pm$ 0.003} & 0.123 $\pm$ 0.003 & \multicolumn{1}{c|}{5.376 $\pm$ 1.174} & \multicolumn{1}{c|}{5.711 $\pm$ 0.962} & 5.527 $ \pm$ 0.803  \\ 
    
            & FedGmTE-Net++ & \multicolumn{1}{c|}{\textcolor{blue}{\textbf{0.120 $\pm$ 0.003}}} & \multicolumn{1}{c|}{\textcolor{blue}{\textbf{0.120 $\pm$ 0.003}}} & \textcolor{blue}{\textbf{0.121 $\pm$ 0.004}} & \multicolumn{1}{c|}{5.234 $\pm$ 1.155} & \multicolumn{1}{c|}{5.163 $\pm$ 1.148} & 5.061 $ \pm$ 1.037  \\
            \hline 
    
            \multirow{4}{*}{$t_{1}$} & NoFedGmTE-Net & \multicolumn{1}{c|}{0.121 $\pm$ 0.004} & \multicolumn{1}{c|}{0.124 $\pm$ 0.005} & 0.120 $\pm$ 0.003 & \multicolumn{1}{c|}{4.962 $\pm$ 1.095}  & \multicolumn{1}{c|}{4.968 $\pm$ 0.962}  & 4.448 $\pm$ 0.794 \\
\hline
        \end{tabular}
        }

    \end{subtable}
    
    \begin{subtable}
        \tablefont
        \centering
        \resizebox{\textwidth}{!}{
        \begin{tabular}{c|c|ccc|ccc}
            & & \multicolumn{3}{c|}{\textcolor{gold}{\textbf{MAE (PC)$\uparrow$}}} & \multicolumn{3}{c}{\textcolor{green}{\textbf{MAE (JD)$\downarrow$}}} \\
            & \multirow[b]{1}{*}{ Methods } & $h_{1}$ & $h_{2}$ & $h_{3}$
            & $h_{1}$ & $h_{2}$ & $h_{3}$ \\
            \hline 
            
            \multirow{4}{*}{$t_{0}$} & NoFedGmTE-Net & \multicolumn{1}{c|}{0.188 $\pm$ 0.065} & \multicolumn{1}{c|}{0.209 $\pm$ 0.049} & 0.250 $\pm$ 0.042 & \multicolumn{1}{c|}{0.776 $\pm$ 0.043}  & \multicolumn{1}{c|}{0.756 $\pm$ 0.019}  & 0.725 $\pm$ 0.014 \\
            
            & FedGmTE-Net & \multicolumn{1}{c|}{0.324 $\pm$ 0.040} & \multicolumn{1}{c|}{0.325 $\pm$ 0.041} &
            0.325 $\pm$ 0.041 & \multicolumn{1}{c|}{0.715 $\pm$ 0.008}  & \multicolumn{1}{c|}{0.714 $\pm$ 0.003}  & 0.717 $\pm$ 0.007 \\
            
            & FedGmTE-Net+ & \multicolumn{1}{c|}{0.509 $\pm$ 0.018} & \multicolumn{1}{c|}{0.511 $\pm$ 0.017} & 0.510 $\pm$ 0.016 & \multicolumn{1}{c|}{0.685 $\pm$ 0.010} & \multicolumn{1}{c|}{0.691 $\pm$ 0.005} & 0.688 $ \pm$ 0.004  \\ 
    
            & FedGmTE-Net++ & \multicolumn{1}{c|}{\textcolor{gold}{\textbf{0.525 $\pm$ 0.012}}} & \multicolumn{1}{c|}{\textcolor{gold}{\textbf{0.525 $\pm$ 0.012}}} & \textcolor{gold}{\textbf{0.525 $\pm$ 0.011}} & \multicolumn{1}{c|}{\textcolor{green}{\textbf{0.672 $\pm$ 0.014}}} & \multicolumn{1}{c|}{\textcolor{green}{\textbf{0.670 $\pm$ 0.012}}} & \textcolor{green}{\textbf{0.668 $ \pm$ 0.015}}  \\
            \hline 
    
            \multirow{4}{*}{$t_{1}$} & NoFedGmTE-Net & \multicolumn{1}{c|}{0.555 $\pm$ 0.059} & \multicolumn{1}{c|}{0.541 $\pm$ 0.044} & 0.584 $\pm$ 0.022 & \multicolumn{1}{c|}{0.640 $\pm$ 0.021}  & \multicolumn{1}{c|}{0.636 $\pm$ 0.010}  & 0.611 $\pm$ 0.020 \\
\hline
        \end{tabular}
        }
    \end{subtable}
    \caption{\textbf{Real Non-IID data}: Functional brain graph trajectory prediction by NoFedGmTE-Net, FedGmTE-Net, FedGmTE-Net+ and FedGmTE-Net++. We highlight in bold, blue for MAE (graph), purple for MAE (NS), gold for MAE (PC), and green for MAE (JD) the best performance at each hospital across timepoints.}
    \label{tab:real_non_iid_func}
\end{table}

\begin{table}
    \tablefont
    \centering
    \begin{subtable}
        \tablefont
        \centering
        \resizebox{\textwidth}{!}{
        \begin{tabular}{c|c|ccc|ccc}
            & & \multicolumn{3}{c|}{\textcolor{blue}{\textbf{MAE (graph)$\downarrow$}}} & \multicolumn{3}{c}{\textcolor{purple}{\textbf{MAE (NS)$\downarrow$}}} \\
            & \multirow[b]{1}{*}{ Methods } & $h_{1}$ & $h_{2}$ & $h_{3}$
            & $h_{1}$ & $h_{2}$ & $h_{3}$ \\
            \hline 
            
            \multirow{4}{*}{$t_{0}$} & NoFedGmTE-Net & \multicolumn{1}{c|}{0.162 $\pm$ 0.017} & \multicolumn{1}{c|}{0.158 $\pm$ 0.007} & 0.170 $\pm$ 0.015 & \multicolumn{1}{c|}{3.046 $\pm$ 0.291}  & \multicolumn{1}{c|}{2.919 $\pm$ 0.985}  & 3.354 $\pm$ 0.312 \\
            
            & FedGmTE-Net & \multicolumn{1}{c|}{0.149 $\pm$ 0.019} & \multicolumn{1}{c|}{0.148 $\pm$ 0.018} &
            0.149 $\pm$ 0.019 & \multicolumn{1}{c|}{2.891 $\pm$ 0.454}  & \multicolumn{1}{c|}{2.821 $\pm$ 0.329}  & 2.845 $\pm$ 0.356 \\
            
            & FedGmTE-Net+ & \multicolumn{1}{c|}{0.133 $\pm$ 0.006} & \multicolumn{1}{c|}{0.133 $\pm$ 0.006} & 0.133 $\pm$ 0.006 & \multicolumn{1}{c|}{2.244 $\pm$ 0.065} & \multicolumn{1}{c|}{\textcolor{purple}{\textbf{2.247 $\pm$ 0.072}}} & 2.261 $ \pm$ 0.068  \\ 
    
            & FedGmTE-Net++ & \multicolumn{1}{c|}{\textcolor{blue}{\textbf{0.129 $\pm$ 0.003}}} & \multicolumn{1}{c|}{\textcolor{blue}{\textbf{0.129 $\pm$ 0.002}}} & \textcolor{blue}{\textbf{0.129 $\pm$ 0.002}} & \multicolumn{1}{c|}{\textcolor{purple}{\textbf{2.192 $\pm$ 0.099}}} & \multicolumn{1}{c|}{2.259 $\pm$ 0.117} & \textcolor{purple}{\textbf{2.258 $ \pm$ 0.124}}  \\
            \hline 
    
            \multirow{4}{*}{$t_{1}$} & NoFedGmTE-Net & \multicolumn{1}{c|}{0.135 $\pm$ 0.008} & \multicolumn{1}{c|}{0.131 $\pm$ 0.003} & 0.136 $\pm$ 0.001 & \multicolumn{1}{c|}{2.289 $\pm$ 0.174}  & \multicolumn{1}{c|}{2.244 $\pm$ 0.099}  & 2.310 $\pm$ 0.111 \\
\hline
        \end{tabular}
        }

    \end{subtable}
    
    \begin{subtable}
        \tablefont
        \centering
        \resizebox{\textwidth}{!}{
        \begin{tabular}{c|c|ccc|ccc}
            & & \multicolumn{3}{c|}{\textcolor{gold}{\textbf{MAE (PC)$\uparrow$}}} & \multicolumn{3}{c}{\textcolor{green}{\textbf{MAE (JD)$\downarrow$}}} \\
            & \multirow[b]{1}{*}{ Methods } & $h_{1}$ & $h_{2}$ & $h_{3}$
            & $h_{1}$ & $h_{2}$ & $h_{3}$ \\
            \hline 
            
            \multirow{4}{*}{$t_{0}$} & NoFedGmTE-Net & \multicolumn{1}{c|}{0.630 $\pm$ 0.070} & \multicolumn{1}{c|}{0.663 $\pm$ 0.030} & 0.583 $\pm$ 0.087 & \multicolumn{1}{c|}{0.412 $\pm$ 0.036}  & \multicolumn{1}{c|}{0.402 $\pm$ 0.010}  & 0.435 $\pm$ 0.033 \\
            
            & FedGmTE-Net & \multicolumn{1}{c|}{0.686 $\pm$ 0.111} & \multicolumn{1}{c|}{0.687 $\pm$ 0.110} &
            0.681 $\pm$ 0.115 & \multicolumn{1}{c|}{0.383 $\pm$ 0.046}  & \multicolumn{1}{c|}{0.381 $\pm$ 0.042}  & 0.384 $\pm$ 0.044 \\
            
            & FedGmTE-Net+ & \multicolumn{1}{c|}{0.774 $\pm$ 0.306} & \multicolumn{1}{c|}{0.775 $\pm$ 0.031} & 0.774 $\pm$ 0.031 & \multicolumn{1}{c|}{0.335 $\pm$ 0.009} & \multicolumn{1}{c|}{0.335 $\pm$ 0.010} & 0.335 $\pm$ 0.010  \\ 
    
            & FedGmTE-Net++ & \multicolumn{1}{c|}{\textcolor{gold}{\textbf{0.789 $\pm$ 0.019}}} & \multicolumn{1}{c|}{\textcolor{gold}{\textbf{0.789 $\pm$ 0.020}}} & \textcolor{gold}{\textbf{0.789 $\pm$ 0.019}} & \multicolumn{1}{c|}{\textcolor{green}{\textbf{0.326 $\pm$ 0.005}}} & \multicolumn{1}{c|}{\textcolor{green}{\textbf{0.329 $\pm$ 0.006}}} & \textcolor{green}{\textbf{0.329 $ \pm$ 0.006}}  \\
            \hline 
    
            \multirow{4}{*}{$t_{1}$} & NoFedGmTE-Net & \multicolumn{1}{c|}{0.801 $\pm$ 0.029} & \multicolumn{1}{c|}{0.825 $\pm$ 0.011} & 0.810 $\pm$ 0.011 & \multicolumn{1}{c|}{0.300 $\pm$ 0.015}  & \multicolumn{1}{c|}{0.291 $\pm$ 0.005}  & 0.307 $\pm$ 0.005 \\
\hline
        \end{tabular}
        }
    \end{subtable}
    \caption{\textbf{Simulated IID data}: Morphological brain graph trajectory prediction by NoFedGmTE-Net, FedGmTE-Net, FedGmTE-Net+ and FedGmTE-Net++. We highlight in bold, blue for MAE (graph), purple for MAE (NS), gold for MAE (PC), and green for MAE (JD) the best performance at each hospital across timepoints.}
    \label{tab:sim_iid_morph}
\end{table}

\begin{table}
    \tablefont
    \centering
    \begin{subtable}
        \tablefont
        \centering
        \resizebox{\textwidth}{!}{
        \begin{tabular}{c|c|ccc|ccc}
            & & \multicolumn{3}{c|}{\textcolor{blue}{\textbf{MAE (graph)$\downarrow$}}} & \multicolumn{3}{c}{\textcolor{purple}{\textbf{MAE (NS)$\downarrow$}}} \\
            & \multirow[b]{1}{*}{ Methods } & $h_{1}$ & $h_{2}$ & $h_{3}$
            & $h_{1}$ & $h_{2}$ & $h_{3}$ \\
            \hline 
            
            \multirow{4}{*}{$t_{0}$} & NoFedGmTE-Net & \multicolumn{1}{c|}{0.122 $\pm$ 0.005} & \multicolumn{1}{c|}{0.121 $\pm$ 0.002} & 0.124 $\pm$ 0.002 & \multicolumn{1}{c|}{\textcolor{purple}{\textbf{3.712 $\pm$ 0.275}}}  & \multicolumn{1}{c|}{4.100 $\pm$ 0.361}  & 4.235 $\pm$ 0.223 \\
            
            & FedGmTE-Net & \multicolumn{1}{c|}{0.118 $\pm$ 0.004} & \multicolumn{1}{c|}{0.118 $\pm$ 0.004} &
            0.118 $\pm$ 0.004 & \multicolumn{1}{c|}{3.790 $\pm$ 0.300}  & \multicolumn{1}{c|}{\textcolor{purple}{\textbf{3.900 $\pm$ 0.300}}}  & \textcolor{purple}{\textbf{3.819 $\pm$ 0.323}} \\
            
            & FedGmTE-Net+ & \multicolumn{1}{c|}{0.108 $\pm$ 0.001} & \multicolumn{1}{c|}{0.108 $\pm$ 0.001} & 0.108 $\pm$ 0.001 & \multicolumn{1}{c|}{5.007 $\pm$ 0.696} & \multicolumn{1}{c|}{4.979 $\pm$ 0.697} & 4.792 $ \pm$ 0.638  \\ 
    
            & FedGmTE-Net++ & \multicolumn{1}{c|}{\textcolor{blue}{\textbf{0.106 $\pm$ 0.008}}} & \multicolumn{1}{c|}{\textcolor{blue}{\textbf{0.106 $\pm$ 0.001}}} & \textcolor{blue}{\textbf{0.106 $\pm$ 0.001}} & \multicolumn{1}{c|}{4.202 $\pm$ 0.299} & \multicolumn{1}{c|}{4.185 $\pm$ 0.238} & 4.207 $ \pm$ 0.231  \\
            \hline 
    
            \multirow{4}{*}{$t_{1}$} & NoFedGmTE-Net & \multicolumn{1}{c|}{0.117 $\pm$ 0.004} & \multicolumn{1}{c|}{0.116 $\pm$ 0.005} & 0.120 $\pm$ 0.005 & \multicolumn{1}{c|}{3.711 $\pm$ 0.858}  & \multicolumn{1}{c|}{3.406 $\pm$ 0.432}  & 3.289 $\pm$ 0.269 \\
\hline
        \end{tabular}
        }

    \end{subtable}
    
    \begin{subtable}
        \tablefont
        \centering
        \resizebox{\textwidth}{!}{
        \begin{tabular}{c|c|ccc|ccc}
            & & \multicolumn{3}{c|}{\textcolor{gold}{\textbf{MAE (PC)$\uparrow$}}} & \multicolumn{3}{c}{\textcolor{green}{\textbf{MAE (JD)$\downarrow$}}} \\
            & \multirow[b]{1}{*}{ Methods } & $h_{1}$ & $h_{2}$ & $h_{3}$
            & $h_{1}$ & $h_{2}$ & $h_{3}$ \\
            \hline 
            
            \multirow{4}{*}{$t_{0}$} & NoFedGmTE-Net & \multicolumn{1}{c|}{0.330 $\pm$ 0.082} & \multicolumn{1}{c|}{0.333 $\pm$ 0.048} & 0.273 $\pm$ 0.044 & \multicolumn{1}{c|}{0.599 $\pm$ 0.017}  & \multicolumn{1}{c|}{0.606 $\pm$ 0.006}  & 0.620 $\pm$ 0.007 \\
            
            & FedGmTE-Net & \multicolumn{1}{c|}{0.402 $\pm$ 0.062} & \multicolumn{1}{c|}{0.406 $\pm$ 0.059} &
            0.400 $\pm$ 0.065 & \multicolumn{1}{c|}{0.587 $\pm$ 0.014}  & \multicolumn{1}{c|}{0.588 $\pm$ 0.014}  & 0.588 $\pm$ 0.014 \\
            
            & FedGmTE-Net+ & \multicolumn{1}{c|}{0.579 $\pm$ 0.005} & \multicolumn{1}{c|}{0.579 $\pm$ 0.005} & 0.578 $\pm$ 0.005 & \multicolumn{1}{c|}{0.575 $\pm$ 0.012} & \multicolumn{1}{c|}{0.574 $\pm$ 0.013} & 0.570 $ \pm$ 0.013  \\ 
    
            & FedGmTE-Net++ & \multicolumn{1}{c|}{\textcolor{gold}{\textbf{0.597 $\pm$ 0.007}}} & \multicolumn{1}{c|}{\textcolor{gold}{\textbf{0.596 $\pm$ 0.006}}} & \textcolor{gold}{\textbf{0.596 $\pm$ 0.006}} & \multicolumn{1}{c|}{\textcolor{green}{\textbf{0.552 $\pm$ 0.008}}} & \multicolumn{1}{c|}{\textcolor{green}{\textbf{0.552 $\pm$ 0.006}}} & \textcolor{green}{\textbf{0.553 $ \pm$ 0.006}}  \\
            \hline 
    
            \multirow{4}{*}{$t_{1}$} & NoFedGmTE-Net & \multicolumn{1}{c|}{0.594 $\pm$ 0.011} & \multicolumn{1}{c|}{0.607 $\pm$ 0.007} & 0.582 $\pm$ 0.023 & \multicolumn{1}{c|}{0.544 $\pm$ 0.019}  & \multicolumn{1}{c|}{0.535 $\pm$ 0.011}  & 0.534 $\pm$ 0.015 \\
\hline
        \end{tabular}
        }
    \end{subtable}
    \caption{\textbf{Simulated IID data}: Functional brain graph trajectory prediction by NoFedGmTE-Net, FedGmTE-Net, FedGmTE-Net+ and FedGmTE-Net++. We highlight in bold, blue for MAE (graph), purple for MAE (NS), gold for MAE (PC), and green for MAE (JD) the best performance at each hospital across timepoints.}
    \label{tab:sim_iid_func}
\end{table}

\begin{table}
    \tablefont
    \centering
    \begin{subtable}
        \tablefont
        \centering
        \resizebox{\textwidth}{!}{
        \begin{tabular}{c|c|ccc|ccc}
            & & \multicolumn{3}{c|}{\textcolor{blue}{\textbf{MAE (graph)$\downarrow$}}} & \multicolumn{3}{c}{\textcolor{purple}{\textbf{MAE (NS)$\downarrow$}}} \\
            & \multirow[b]{1}{*}{ Methods } & $h_{1}$ & $h_{2}$ & $h_{3}$
            & $h_{1}$ & $h_{2}$ & $h_{3}$ \\
            \hline 
            
            \multirow{4}{*}{$t_{0}$} & NoFedGmTE-Net & \multicolumn{1}{c|}{0.168 $\pm$ 0.008} & \multicolumn{1}{c|}{0.169 $\pm$ 0.020} & 0.151 $\pm$ 0.002 & \multicolumn{1}{c|}{3.212 $\pm$ 0.205}  & \multicolumn{1}{c|}{3.136 $\pm$ 0.290}  & 3.055 $\pm$ 0.119 \\
            
            & FedGmTE-Net & \multicolumn{1}{c|}{0.156 $\pm$ 0.009} & \multicolumn{1}{c|}{0.158 $\pm$ 0.011} &
            0.157 $\pm$ 0.008 & \multicolumn{1}{c|}{2.822 $\pm$ 0.215}  & \multicolumn{1}{c|}{3.177 $\pm$ 0.447}  & 3.003 $\pm$ 0.229 \\
            
            & FedGmTE-Net+ & \multicolumn{1}{c|}{0.138 $\pm$ 0.002} & \multicolumn{1}{c|}{0.139 $\pm$ 0.002} & 0.138 $\pm$ 0.002 & \multicolumn{1}{c|}{2.304 $\pm$ 0.083} & \multicolumn{1}{c|}{\textcolor{purple}{\textbf{2.302 $\pm$ 0.223}}} & \textcolor{purple}{\textbf{2.307 $ \pm$ 0.132}}  \\ 
    
            & FedGmTE-Net++ & \multicolumn{1}{c|}{\textcolor{blue}{\textbf{0.133 $\pm$ 0.004}}} & \multicolumn{1}{c|}{\textcolor{blue}{\textbf{0.134 $\pm$ 0.004}}} & \textcolor{blue}{\textbf{0.134 $\pm$ 0.004}} & \multicolumn{1}{c|}{\textcolor{purple}{\textbf{2.209 $\pm$ 0.081}}} & \multicolumn{1}{c|}{2.357 $\pm$ 0.224} & 2.328 $ \pm$ 0.149  \\
            \hline 
    
            \multirow{4}{*}{$t_{1}$} & NoFedGmTE-Net & \multicolumn{1}{c|}{0.139 $\pm$ 0.005} & \multicolumn{1}{c|}{0.137 $\pm$ 0.005} & 0.136 $\pm$ 0.004 & \multicolumn{1}{c|}{2.428 $\pm$ 0.063}  & \multicolumn{1}{c|}{2.328 $\pm$ 0.121}  &  2.434 $\pm$ 0.286 \\
\hline
        \end{tabular}
        }

    \end{subtable}
    
    \begin{subtable}
        \tablefont
        \centering
        \resizebox{\textwidth}{!}{
        \begin{tabular}{c|c|ccc|ccc}
            & & \multicolumn{3}{c|}{\textcolor{gold}{\textbf{MAE (PC)$\uparrow$}}} & \multicolumn{3}{c}{\textcolor{green}{\textbf{MAE (JD)$\downarrow$}}} \\
            & \multirow[b]{1}{*}{ Methods } & $h_{1}$ & $h_{2}$ & $h_{3}$
            & $h_{1}$ & $h_{2}$ & $h_{3}$ \\
            \hline 
            
            \multirow{4}{*}{$t_{0}$} & NoFedGmTE-Net & \multicolumn{1}{c|}{0.586 $\pm$ 0.049} & \multicolumn{1}{c|}{0.608 $\pm$ 0.087} & 0.685 $\pm$ 0.029 & \multicolumn{1}{c|}{0.429 $\pm$ 0.019}  & \multicolumn{1}{c|}{0.428 $\pm$ 0.041}  & 0.391 $\pm$ 0.010 \\
            
            & FedGmTE-Net & \multicolumn{1}{c|}{0.650 $\pm$ 0.070} & \multicolumn{1}{c|}{0.651 $\pm$ 0.071} &
            0.650 $\pm$ 0.068 & \multicolumn{1}{c|}{0.397 $\pm$ 0.026}  & \multicolumn{1}{c|}{0.409 $\pm$ 0.033}  & 0.404 $\pm$ 0.024 \\
            
            & FedGmTE-Net+ & \multicolumn{1}{c|}{0.749 $\pm$ 0.008} & \multicolumn{1}{c|}{0.746 $\pm$ 0.009} & 0.747 $\pm$ 0.008 & \multicolumn{1}{c|}{0.349 $\pm$ 0.007} & \multicolumn{1}{c|}{0.350 $\pm$ 0.007} & 0.349 $\pm$ 0.006  \\ 
    
            & FedGmTE-Net++ & \multicolumn{1}{c|}{\textcolor{gold}{\textbf{0.750 $\pm$ 0.032}}} & \multicolumn{1}{c|}{\textcolor{gold}{\textbf{0.752 $\pm$ 0.031}}} & \textcolor{gold}{\textbf{0.752 $\pm$ 0.032}} & \multicolumn{1}{c|}{\textcolor{green}{\textbf{0.336 $\pm$ 0.008}}} & \multicolumn{1}{c|}{\textcolor{green}{\textbf{0.341 $\pm$ 0.011}}} & \textcolor{green}{\textbf{0.339 $ \pm$ 0.009}}  \\
            \hline 
    
            \multirow{4}{*}{$t_{1}$} & NoFedGmTE-Net & \multicolumn{1}{c|}{0.786 $\pm$ 0.011} & \multicolumn{1}{c|}{0.811 $\pm$ 0.009} & 0.806 $\pm$ 0.023 & \multicolumn{1}{c|}{0.308 $\pm$ 0.004}  & \multicolumn{1}{c|}{0.305 $\pm$ 0.004}  & 0.311 $\pm$ 0.008 \\
\hline
        \end{tabular}
        }
    \end{subtable}
    \caption{\textbf{Simulated Non-IID data}: Morphological brain graph trajectory prediction by NoFedGmTE-Net, FedGmTE-Net, FedGmTE-Net+ and FedGmTE-Net++. We highlight in bold, blue for MAE (graph), purple for MAE (NS), gold for MAE (PC), and green for MAE (JD) the best performance at each hospital across timepoints.}
    \label{tab:sim_non_iid_morph}
\end{table}

\begin{table}[ht!]
    \tablefont
    \centering
    \begin{subtable}
        \tablefont
        \centering
        \resizebox{\textwidth}{!}{
        \begin{tabular}{c|c|ccc|ccc}
            & & \multicolumn{3}{c|}{\textcolor{blue}{\textbf{MAE (graph)$\downarrow$}}} & \multicolumn{3}{c}{\textcolor{purple}{\textbf{MAE (NS)$\downarrow$}}} \\
            & \multirow[b]{1}{*}{ Methods } & $h_{1}$ & $h_{2}$ & $h_{3}$
            & $h_{1}$ & $h_{2}$ & $h_{3}$ \\
            \hline 
            
            \multirow{4}{*}{$t_{0}$} & NoFedGmTE-Net & \multicolumn{1}{c|}{0.126 $\pm$ 0.002} & \multicolumn{1}{c|}{0.125 $\pm$ 0.006} & 0.119 $\pm$ 0.003 & \multicolumn{1}{c|}{3.984 $\pm$ 0.883}  & \multicolumn{1}{c|}{\textcolor{purple}{\textbf{3.922 $\pm$ 0.555}}}  & 4.277 $\pm$ 0.604 \\
            
            & FedGmTE-Net & \multicolumn{1}{c|}{0.122 $\pm$ 0.004} & \multicolumn{1}{c|}{0.121 $\pm$ 0.004} &
            0.121 $\pm$ 0.004 & \multicolumn{1}{c|}{\textcolor{purple}{\textbf{3.787 $\pm$ 0.397}}}  & \multicolumn{1}{c|}{4.091 $\pm$ 0.495}  & \textcolor{purple}{\textbf{4.012 $\pm$ 0.520}} \\
            
            & FedGmTE-Net+ & \multicolumn{1}{c|}{0.109 $\pm$ 0.001} & \multicolumn{1}{c|}{0.109 $\pm$ 0.001} & 0.109 $\pm$ 0.001 & \multicolumn{1}{c|}{4.829 $\pm$ 0.653} & \multicolumn{1}{c|}{4.752 $\pm$ 0.219} & 4.765 $ \pm$ 0.590  \\ 
    
            & FedGmTE-Net++ & \multicolumn{1}{c|}{\textcolor{blue}{\textbf{0.108 $\pm$ 0.009}}} & \multicolumn{1}{c|}{\textcolor{blue}{\textbf{0.107 $\pm$ 0.001}}} & \textcolor{blue}{\textbf{0.107 $\pm$ 0.001}} & \multicolumn{1}{c|}{3.847 $\pm$ 0.484} & \multicolumn{1}{c|}{4.234 $\pm$ 0.723} & 4.103 $ \pm$ 0.464  \\
            \hline 
    
            \multirow{4}{*}{$t_{1}$} & NoFedGmTE-Net & \multicolumn{1}{c|}{0.119 $\pm$ 0.005} & \multicolumn{1}{c|}{0.117 $\pm$ 0.001} & 0.116 $\pm$ 0.003 & \multicolumn{1}{c|}{3.489 $\pm$ 0.196}  & \multicolumn{1}{c|}{\textcolor{purple}{\textbf{3.767 $\pm$ 0.568}}}  & 3.880 $\pm$ 0.752 \\
\hline
        \end{tabular}
        }

    \end{subtable}
    
    \begin{subtable}
        \tablefont
        \centering
        \resizebox{\textwidth}{!}{
        \begin{tabular}{c|c|ccc|ccc}
            & & \multicolumn{3}{c|}{\textcolor{gold}{\textbf{MAE (PC)$\uparrow$}}} & \multicolumn{3}{c}{\textcolor{green}{\textbf{MAE (JD)$\downarrow$}}} \\
            & \multirow[b]{1}{*}{ Methods } & $h_{1}$ & $h_{2}$ & $h_{3}$
            & $h_{1}$ & $h_{2}$ & $h_{3}$ \\
            \hline 
            
            \multirow{4}{*}{$t_{0}$} & NoFedGmTE-Net & \multicolumn{1}{c|}{0.289 $\pm$ 0.031} & \multicolumn{1}{c|}{0.300 $\pm$ 0.102} & 0.377 $\pm$ 0.038 & \multicolumn{1}{c|}{0.613 $\pm$ 0.016}  & \multicolumn{1}{c|}{0.611 $\pm$ 0.016}  & 0.602 $\pm$ 0.018 \\
            
            & FedGmTE-Net & \multicolumn{1}{c|}{0.350 $\pm$ 0.057} & \multicolumn{1}{c|}{0.349 $\pm$ 0.061} &
            0.350 $\pm$ 0.054 & \multicolumn{1}{c|}{0.595 $\pm$ 0.018}  & \multicolumn{1}{c|}{0.603 $\pm$ 0.018}  & 0.600 $\pm$ 0.018 \\
            
            & FedGmTE-Net+ & \multicolumn{1}{c|}{0.558 $\pm$ 0.012} & \multicolumn{1}{c|}{0.557 $\pm$ 0.011} & 0.557 $\pm$ 0.010 & \multicolumn{1}{c|}{0.576 $\pm$ 0.010} & \multicolumn{1}{c|}{0.575 $\pm$ 0.003} & 0.575 $ \pm$ 0.009  \\ 
    
            & FedGmTE-Net++ & \multicolumn{1}{c|}{\textcolor{gold}{\textbf{0.578 $\pm$ 0.008}}} & \multicolumn{1}{c|}{\textcolor{gold}{\textbf{0.580 $\pm$ 0.008}}} & \textcolor{gold}{\textbf{0.579 $\pm$ 0.009}} & \multicolumn{1}{c|}{\textcolor{green}{\textbf{0.547 $\pm$ 0.014}}} & \multicolumn{1}{c|}{\textcolor{green}{\textbf{0.556 $\pm$ 0.018}}} & \textcolor{green}{\textbf{0.554 $\pm$ 0.011}}  \\
            \hline 
    
            \multirow{4}{*}{$t_{1}$} & NoFedGmTE-Net & \multicolumn{1}{c|}{0.590 $\pm$ 0.013} & \multicolumn{1}{c|}{0.701 $\pm$ 0.006} & 0.594 $\pm$ 0.016 & \multicolumn{1}{c|}{0.543 $\pm$ 0.010}  & \multicolumn{1}{c|}{0.549 $\pm$ 0.014}  & 0.546 $\pm$ 0.018 \\
\hline
        \end{tabular}
        }
    \end{subtable}
    \caption{\textbf{Simulated Non-IID data}: Functional brain graph trajectory prediction by NoFedGmTE-Net, FedGmTE-Net, FedGmTE-Net+ and FedGmTE-Net++. We highlight in bold, blue for MAE (graph), purple for MAE (NS), gold for MAE (PC), and green for MAE (JD) the best performance at each hospital across timepoints.}
    \label{tab:sim_non_iid_func}
\end{table}

Finally, to further validate our framework and substantiate its effectiveness within a more extensive dataset environment, we carried out parallel experiments for both IID and non-IID scenarios utilizing our simulated dataset, which comprises 100 distinct subjects, with each hospital contributing 25 subjects. The outcomes of these experiments are presented in \textbf{Table \ref{tab:sim_iid_morph}} and \textbf{Table \ref{tab:sim_iid_func}} for the IID case and in \textbf{Table \ref{tab:sim_non_iid_morph}}  and \textbf{Table \ref{tab:sim_non_iid_func}} for the non-IID cases. 
It is crucial to note that this dataset surpasses the scale of the real dataset we initially examined. Our research, however, remains centred on addressing data limitations, as all our proposed methods and their variants are specifically designed to confront scenarios marked by data scarcity. Despite the larger dataset employed in these experiments, the results maintain a good quality. Similarly to the real dataset, for the IID case, FedGmTE-Net++ overall outperforms the alternatives in most scenarios for the MAE(graph), MAE(PCC), and MAE(JD) measures, while its performance for the MAE(NS) tends to exhibit more randomness. In the non-IID scenario, as anticipated, our proposed method exhibits a slight decrease in performance. However, it still emerges as the top-performing approach across most evaluation measures. The above highlights the reliability of our proposed methods which performs well even in the larger data scenarios.

\subsection{Brain graph predictions visualization}

\begin{figure}[ht!]
\centering
{\includegraphics[width=11cm]
{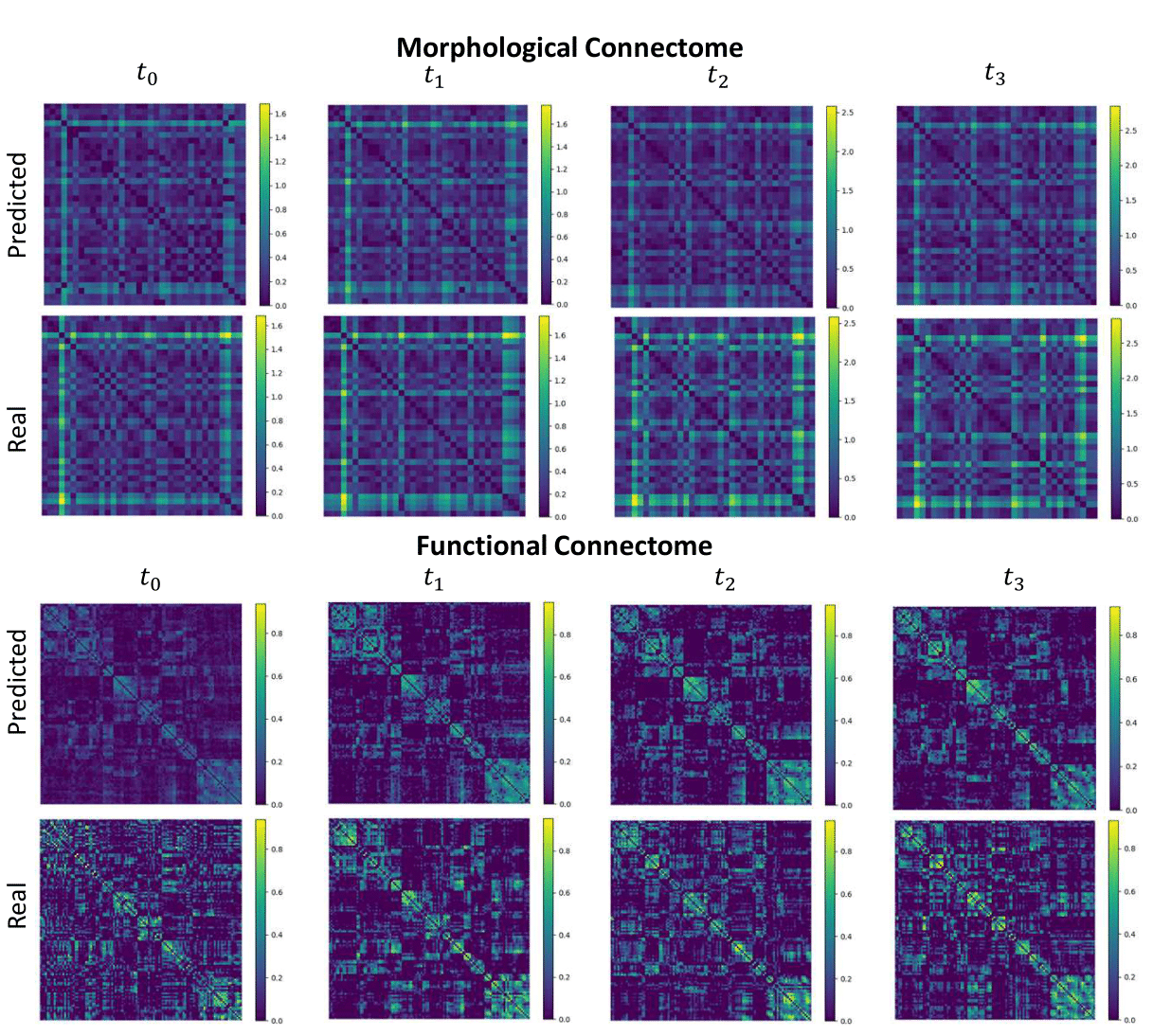}}
\caption{ Predicted against real for morphological and functional connectomes at $t_{0}$, $t_{1}$, $t_{2}$ and $t_{3}$ by FedGmTE-Net++ of a representative subject.}
\label{fig:vis}
\end{figure}

To better showcase the effectiveness of our network, we present visual comparisons of the real brain graphs and their corresponding predictions for both morphological and functional graph evolution trajectories in \textbf{Fig.} \ref{fig:vis}. These visualizations demonstrate the strong performance of our framework, especially considering the challenging scenario and our simplistic encoder-decoder architectures. Despite utilizing only a single modality input (morphological since it is more affordable) at the baseline timepoint along with a very limited dataset, we achieve promising multimodal graph trajectory evolution prediction across time. 


\section{Discussion}

In this work, we introduce FedGmTE-Net++, the first graph multi-trajectory framework that leverages federation for forecasting the infant brain evolution trajectory in a data-scarce environment. The emphasis of the framework is on forecasting the dynamic progression of brain development across multiple timepoints and imaging modalities, all while relying on a single baseline brain graph as input.

FedGmTE-Net constitutes the first federated learning framework specifically designed for predicting graph brain multi-trajectory evolution. By leveraging federated learning, we combine the knowledge gained from diverse hospitals with small local training datasets and hence significantly improve the predictive performance of each hospital's GNN model. We further ensure that the distinctive topological features within the brain graphs are preserved by incorporating an additional topology loss component to the model's objective function. Maintaining the rich array of topological patterns inherent in the brain connectome can play a pivotal role in capturing a more comprehensive prediction of its multi-facet connectivity.

With FedGmTE-Net+, we further introduce a novel auxiliary regularizer that utilizes the entirety of available temporal graphs to substantially enhance the initial baseline model's predictions. Consequently, this creates a chain reaction that ultimately results in performance improvements across all subsequent timepoints. Additionally, we present an efficient approach to tackle the challenge of missing data, involving a KNN-based imputation technique as part of the data preprocessing phase. By utilizing the baseline imaging graphs, accessible for all subjects, we fill in the missing graphs along the evolution trajectories at subsequent timepoints. This proactive imputation strategy prevents the loss of valuable information from incomplete subjects.

For our final and optimal framework, FedGmTE-Net++, we build on the limitations of the previous variant and propose a refinement process in the final stages of the training. This refinement step improves on the initial naive imputations generated during the precompletion, thereby elevating the dataset's overall quality and enabling a more efficient utilization of incomplete subjects. Subsequently, the final dataset is employed for fine-tuning the entire network, yielding substantial improvements in overall performance. 

In our extensive experiments, we have initially showcased the practicality and benefits of federated learning as a collaborative framework for hospitals, especially in situations where data constraints are a primary concern. Moreover, we thoroughly assessed and compared our proposed method across benchmarks, illustrating its performance across four distinct evaluation measures and in four diverse data scenarios. These scenarios encompassed both independent and non-independent data cases, as well as real-world and simulated datasets. Notably, FedGmTe-Net+ introduces a valuable auxiliary regularization, leading to a notable enhancement in performance at the baseline timepoint, triggering a sequence of smaller improvements in subsequent timepoints. While, FedGmTE-Net++, with its emphasis on imputation refinement, consistently delivers overall improvements across all timepoints and outperforms its predecessors.

Our proposed framework also stands out as a robust, versatile, and easily adaptable framework. Its various modules are highly customizable, thanks to its flexible architecture. A minor adjustment, such as modifying network layers and fine-tuning different hyperparameters, is all that is required to seamlessly transition and operate to a different type of network or scenario. In our experiments, when presented with a ($35 \times 35$) dimensional input brain graph, our framework generates two trajectories, each comprising a series of subsequent brain graphs specific to a particular modality. Our focus lies on the morphological and functional connectomes. For the morphological connectome, the matrices have a size of ($35 \times 35$), while for the functional connectome, they extend to a size of ($116 \times 116$). However, there is flexibility to choose alternative trajectories and imaging modalities for the experiments. Hence, various and different imaging techniques can be opted for (i.e., Voxel-Based Morphometry (VBM) \citep{ashburner2000voxel}, Diffusion Tensor Imaging \citep{jones2011diffusion, basser1994mr}). In that way, our framework is generic and can be used for any type of isomorphic graph. Our model can also extend beyond its application in brain evolution predictive tasks, encompassing various areas of research. It can be employed in any scenario represented as a graph, aligning with our generic longitudinal multi-trajectory framework that leverages GNNs. For example, it can be applied in social network analysis \citep{otte2002social}, depicting the evolution of organizational hierarchies, or in the realm of molecular dynamics within chemistry \citep{xie2006molecular}, where it can model the dynamic changes occurring in molecular structures over time.
 
\textbf{Limitations and future directions.}
Although our proposed method significantly outperformed its benchmarks and variants, it has a few limitations that can be addressed in future work. First, we opted for the use of GCNs as the foundational blocks of our network, which are quite simplistic. However, in our future work, we would like to consider alternative architectures and mechanisms like graph attention \citep{velivckovic2017graph} or GRNs \citep{huang2019graph}. Their use can facilitate the preservation and communication of memory across distant timepoints, allowing for long-term dependencies to be maintained. Second, we generated our multi-trajectory brain predictions by leveraging two distinct MR imaging modalities (T1-w MRI and fMRI), with the T1-w modality serving as the input. It would be interesting to also evaluate our FedGmTE-Net++ on brain trajectories derived from and across different neuroimaging modalities such as diffusion MRIs. Third,  we aspire to investigate additional avenues within the realm of federated learning beyond the conventional FedAvg \citep{mcmahan2017communication} algorithm. For example, \citep{zhang2022proportional} is an interesting paper that can inspire the work in fairness among hospital updates. Alternatively, addressing security concerns, such as countering poison attacks \citep{alkhunaizi2022suppressing}, or ensuring robustness against malicious clients \citep{cao2021provably}, presents an intriguing research path as well.

\section{Conclusion}
In conclusion, in this work, we present FedGmTE-Net++, a pioneering federated learning framework specifically designed for predicting the infant brain graph \emph{multi-trajectory} evolution from a single modality graph in \emph{data-scarce environments}. By combining the unique knowledge from diverse hospitals with small local training datasets, our approach substantially enhances the predictive power of each hospital's GNN model with the help of federated learning. Additionally, we introduce an innovative auxiliary regularizer that leverages all available temporal graphs, and thus significantly enhances the accuracy of the initial baseline model. Ultimately this leads to performance improvements across all subsequent timepoints and imaging modalities. Lastly, we introduce a supervised refined imputation technique that utilizes trained similarity regressors to elevate the quality of missing brain graph predictions at the later stages of training, further fine-tuning our models for improved performance. In our future work, we will extend Fed-GmTE-Net++ to accommodate the diverse hardware and software capabilities presented in highly heterogeneous hospital environments.

\section*{Competing Interests}
The authors declare no competing interests.

\newpage
\bibliography{Biblio3}
\bibliographystyle{model2-names}


\end{document}